\documentclass[twocolumn,floatfix,showpacs,tightenlines,superscriptaddress]{revtex4-2}
\usepackage{mathtools}
\usepackage{bm}
\usepackage{dsfont,amsthm,amsbsy}
\usepackage{verbatim}
\usepackage{amssymb}
\usepackage{amsmath}
\usepackage{bbm}
\usepackage{graphicx}
\usepackage{epstopdf}
\usepackage{subfigure}
\usepackage{natbib}
\usepackage{epsfig}
\usepackage{amsfonts}
\usepackage{mathrsfs}
\usepackage{sidecap}
\usepackage{lipsum}
\usepackage[toc,page,title,titletoc,header]{appendix}
\usepackage[colorlinks,linkcolor=blue,citecolor=blue,anchorcolor=blue,urlcolor=blue]{hyperref}
\usepackage{hyperref}
\usepackage{resizegather}
\usepackage{tikz}
\usepackage{float}
\usepackage{mathbbol}
\usepackage[normalem]{ulem}
\usepackage{cancel}
\usepackage{upgreek}

\usepackage{dblfloatfix}
\usepackage{graphicx}  
\usepackage{dcolumn}   
\usepackage{bm}        
\usepackage{amssymb}   
\usepackage{amsmath}
\usepackage{lipsum}
\usepackage{xcolor}
\usepackage{ulem}
\usepackage{braket}
\def\i{\text {Im}}
\def\r{\text {Re}}

\begin{document}

\title{Dynamical susceptibility and quantum Fisher information in the Su--Schrieffer--Heeger model with Hatsugai--Kohmoto interactions}
\author{Sepide Mohamadi}
\affiliation{Department of Physics, Institute for Advanced Studies in Basics Sciences (IASBS), Zanjan 45137- 66731, Iran}
\author{Jahanfar Abouie}
\email{jahan@iasbs.ac.ir}
\affiliation{Department of Physics, Institute for Advanced Studies in Basics Sciences (IASBS), Zanjan 45137- 66731, Iran}

\date{\today}

\begin{abstract}
We investigate the dynamical spin and charge susceptibilities and the associated quantum Fisher information in a class of interacting lattice models, with a primary focus on the Su--Schrieffer--Heeger model in the presence of Hatsugai--Kohmoto interactions. To provide a rigorous analytical benchmark, we contrast the response properties of the SSH--HK system with those of the single-band Hubbard and SSH--Hubbard models, treated within the random-phase approximation, and incorporate the strong-coupling limit through the Müller susceptibility. While standard Hubbard-type interactions typically suppress excitation strength, we demonstrate that the SSH--HK model displays qualitatively distinct physical behavior arising from the interplay between SSH dimerization and the momentum-diagonal nature of the HK interaction. 
Leveraging the exact solvability of the HK term, we derive closed-form expressions for the dynamical susceptibility, revealing unique filling-controlled characteristics such as a finite response at zero wave vector and a pronounced restructuring of spectral weight across integer and fractional filling sectors. We show that the quantum Fisher information--defined as the frequency integral of the imaginary part of the susceptibility--serves as an efficient probe of these filling sectors, exhibiting distinct piecewise behavior that distinguishes integer from fractional fillings. Notably, our results indicate that the quantum Fisher information remains insensitive to topological transitions within uniform-density regimes, highlighting the limitations of standard dynamical response functions in characterizing band topology. These findings establish the SSH--HK model as a powerful analytical platform for exploring the competition between topology and strong correlations, demonstrating how dynamical susceptibilities and the quantum Fisher information provide complementary, experimentally accessible probes of many-body physics.
\end{abstract}

\maketitle
\section{Introduction}
Linear-response theory provides a unifying framework for characterizing how quantum systems react to weak external perturbations, with applications ranging from condensed-matter physics to optical spectroscopy, materials science, and quantum-information processing \cite{bloch2008many,fetter2012quantum,rech2015}. In particular, dynamical charge and spin susceptibilities, especially their imaginary parts, encode the spectrum of elementary excitations and the underlying many-body correlations \cite{tohyama1994exact,preuss1997pseudogaps,bickers1989conserving,charfi1992spin,dahm1995quasiparticle,tohyama1995spin,lohneysen2007fermi}. They characterize the dispersion and damping of collective modes such as paramagnons \cite{lebert2023paramagnon} and charge-density-wave excitations \cite{coleman2015introduction,mahan2013many,bruus2004many,gruner1988dynamics}, and capture critical scaling near quantum phase transitions \cite{millis1993effect}. Through the fluctuation–dissipation theorem and sum rules, these response functions also connect spectroscopic observables to entanglement-sensitive quantities, including the quantum Fisher information (QFI) \cite{hauke2016measuring,abouie2024entanglement,greif2013}. They therefore provide a unified framework for relating nonequilibrium dynamics \cite{konic2019,pol2011}, topology, and quantum information in correlated systems.

The Su--Schrieffer--Heeger (SSH) chain is a prototypical one-dimensional model exhibiting topological insulating phases \cite{su1979solitons,ahmadi2020topological}. Its topological properties have been widely investigated theoretically and demonstrated experimentally in several platforms, including cold-atom optical lattices \cite{atala2013direct,st2017lasing,blanco2016topological}, photonic waveguide arrays \cite{rechtsman2013photonic}, and acoustic metamaterials \cite{xiao2015synthetic}. The model describes a dimerized one-dimensional lattice with two sublattices, $a$ and $b$, per unit cell and alternating hopping amplitudes $v$ and $w$. Its Hamiltonian is
\begin{equation}
\label{s1.1}
H_{\mathrm{SSH}}
=
\sum_{j,\sigma}
\left(
v\, a_{j\sigma}^\dagger b_{j\sigma}
+
w\, a_{j+1,\sigma}^\dagger b_{j\sigma}
+ \mathrm{H.c.}
\right),
\end{equation}
where $a_{j\sigma}^\dagger$ ($a_{j\sigma}$) and $b_{j\sigma}^\dagger$ ($b_{j\sigma}$) create (annihilate) a fermion with spin $\sigma$ on the $a$ and $b$ sublattices of the $j$th unit cell, respectively. The parameters $v$ and $w$ denote the intracell and intercell hopping amplitudes. The model undergoes a topological phase transition at $v=w$, which separates a trivial insulating phase for $v>w$ from a topological phase for $v<w$. In the latter case, the system supports topologically protected edge states when open boundary conditions are imposed.

Interacting extensions of topological models, such as the SSH--Hubbard chain, have recently attracted considerable interest \cite{wu2006helical,zheng2011particle,hohenadler2011correlation,yu2011mott}. On-site repulsive interactions generate competing tendencies toward bond-order formation \cite{feng2022phase}, charge ordering \cite{xing2023attractive}, and enhanced antiferromagnetic spin correlations \cite{urasaki1998thermal}.
 Even richer behavior arises in the presence of long-range Hatsugai-Kohmoto (HK) interactions \cite{hatsugai1992exactly}, which produce a ground-state phase diagram containing both topological and trivial non-Fermi-liquid (NFL) phases \cite{mohamadi2025emergence}. Such phases lie beyond the conventional Fermi-liquid paradigm and are characterized by the absence of long-lived quasiparticles. In particular, the interacting SSH chain has been shown to host a topological non-Fermi-liquid (TNFL) phase with unconventional quasiparticle properties and finite electric polarization \cite{mohamadi2025emergence}. Despite these advances, the dynamical properties of the TNFL phase remain largely unexplored.

In this work, we investigate the linear-response properties of the SSH--HK model across a broad range of interaction regimes. We compute the dynamical charge and spin susceptibilities for the exactly solvable SSH--HK model, as well as for the non-exactly-solvable Hubbard and SSH--Hubbard models. For the solvable cases, we derive exact analytical expressions for the dynamical susceptibilities in different phases and for external fields with various wave vectors. We show that the interplay between lattice dimerization and long-range HK interactions produces dynamical features absent in the SSH--Hubbard model. In particular, the SSH--HK model exhibits a finite response to spatially uniform external fields at zero wave vector.

By integrating the imaginary parts of dynamical susceptibilities over frequency, we extract the QFI, which provides a robust and experimentally accessible witness of multipartite entanglement \cite{strobel2014fisher,hyllus2012fisher}. We show that the QFI is a sensitive probe of phases at different fillings and of interaction-driven transitions. Because it can be obtained directly from measurable response functions, it offers a practical route for characterizing correlated quantum phases in experiments, including ultracold atoms, neutron scattering, and pump-probe spectroscopy \cite{hauke2016measuring}.

The remainder of this paper is organized as follows. In Sec.~\ref{sec:notsolvable}, we analyze the dynamical susceptibility of the Hubbard model, and in Sec.~\ref{sec:notsolvablesshh} we extend the discussion to the SSH--Hubbard model. We then turn to exactly solvable cases, discussing the HK model in Sec.~\ref{sec:solvable} and the SSH--HK model in Sec.~\ref{sec:solvablesshhk}. In Sec.~\ref{sec:qfi}, we examine the QFI and its relation to multipartite entanglement, emphasizing its utility in distinguishing the different phases of the system.

\section{Dynamical susceptibility of interacting systems}

Within linear-response theory, the effect of a time-dependent magnetic field on a quantum system is characterized by the dynamical spin susceptibility. This quantity can be directly probed in spectroscopic experiments, making it a natural starting point for our analysis. The spin and charge susceptibilities are defined in terms of transverse and longitudinal spin correlation functions as \cite{mahan2013many}
\begin{eqnarray}\label{eq:chi}
\chi_s(q,\tau) &=& \chi^{+-}(q, \tau) = - \left\langle T_\tau s^{+}(q, \tau) s^{-}(-q, 0)\right\rangle ,\\
\chi_c(q,\tau) &=& \chi^{zz}(q, \tau) = - \left\langle T_\tau s^{z}(q, \tau) s^{z}(-q, 0)\right\rangle ,
\end{eqnarray}
where $s^{\pm}(q, \tau)= s^{x}(q, \tau) \pm i s^{y}(q, \tau)$ and $s^{z}(q, \tau)$ denote components of the spin operator at wave vector $q$ and imaginary time $\tau$. The wave vector $q$ characterizes the spatial periodicity of the external perturbation and corresponds to the momentum carried by the external field, while $\tau$ denotes the imaginary time. Here $T_\tau$ is the imaginary-time ordering operator and $\langle \cdots \rangle$ represents the thermal average. The minus sign in Eqs.~(\ref{eq:chi}) follows from the standard Matsubara formulation of linear-response theory and ensures the correct analytic continuation to the retarded susceptibility as well as a positive spectral weight. The appearance of $-q$ in the second operator follows from the Fourier transformation of the real-space spin-spin correlator and guarantees momentum conservation in a translationally invariant system. Physically, $s^{+}(q)$ creates a spin excitation with momentum $q$, whereas $s^{-}(-q)$ annihilates an excitation with momentum $-q$, so that the total momentum of the response function is conserved.

The spin operators can be expressed in terms of electron creation and annihilation operators as
\begin{eqnarray}
s^{\alpha}(q, \tau)= \frac{1}{N} \sum_{k \sigma \sigma^\prime} 
c^\dagger_{k+q,\sigma}(\tau) 
\frac{\sigma^\alpha_{\sigma \sigma^\prime}}{2} 
c_{k,\sigma^\prime}(\tau),
\end{eqnarray}
where $\sigma^\alpha$ with $\alpha=x, y, z$ are Pauli matrices. 
Using this representation, the spin and charge susceptibilities can be written in terms of fermionic operators as
\begin{eqnarray} \label{eq5.1.1}
\nonumber&&\chi_s(q,\tau)= \\
&&-\frac{1}{N} \sum_{k k^\prime}
\left\langle T_{\tau}
c^{\dagger}_{k+q,\uparrow} (\tau)
c_{k,\downarrow}(\tau)
c^{\dagger}_{k^\prime-q,\downarrow}(0)
c_{k^\prime,\uparrow }(0)
\right\rangle,\\ \label{eq5.1.2}
\nonumber&&\chi_c(q,\tau)=\\
&&-\frac{1}{N} \sum_{k k^\prime \sigma \sigma^\prime}
\left\langle T_{\tau}
c^{\dagger}_{k+q, \sigma}(\tau)
c_{k, \sigma}(\tau)
c^{\dagger}_{k^\prime-q, \sigma^\prime}(0)
c_{k^\prime, \sigma^\prime}(0)
\right\rangle.
\end{eqnarray}

For non-interacting free fermion systems, the four-operator expectation values in Eqs.~\eqref{eq5.1.1} and \eqref{eq5.1.2} can be evaluated using Wick’s theorem, reducing them to products of single-particle Green’s functions. This yields
\begin{align}\label{eq:nonint-sus1}
    \chi_s^{0}(q,\tau) = \frac{1}{N} \sum_{k,\sigma} 
    G^{0}_\sigma(k+q,-\tau)\, G^{0}_\sigma(k,\tau),
\end{align}
where $G^{0}_\sigma(k,\tau)=-\langle T_\tau c_{k\sigma}(\tau)\, c^\dagger_{k\sigma}(0)\rangle$
is the single-particle Green’s function for spin $\sigma$, and the superscript $0$ denotes the non-interacting case. The absence of an overall minus sign in Eq.~\eqref{eq:nonint-sus1} reflects the cancellation between the prefactor in the susceptibility definition and the fermionic sign generated by Wick contractions, together with the minus signs already contained in the single-particle Green’s functions.

In this non-interacting, spin-symmetric model—where spin-up and spin-down fermions have equal densities—the charge susceptibility is directly related to the spin susceptibility. In particular, for such systems the charge susceptibility equals twice the longitudinal spin susceptibility. Since all models considered in this work satisfy this condition, we focus on the spin susceptibility and therefore omit the subscript $s$ in what follows.

For free fermion systems, the single-particle Green’s function in frequency space takes the form
\[
G^0_\sigma(k,\omega)=\frac{1}{\omega+i\eta-\xi_k},
\]
where \(\xi_k=\varepsilon_k-\mu\). Using this expression, the dynamical susceptibility as a function of the external field wavevector \(q\) and frequency \(\omega\) becomes
\begin{align}\label{eq:nonint-sus}
    \chi^{0}(q,\omega)
    =
    \frac{2}{N}\sum_k
    \frac{\Theta(-\xi_k)-\Theta(-\xi_{k+q})}
    {\omega+i\eta+\varepsilon_k-\varepsilon_{k+q}},
\end{align}
where \(\varepsilon_k\) is the non-interacting dispersion, \(\mu\) is the chemical potential, and \(\Theta\) denotes the Heaviside step function. Here, $\Theta(-\xi_k)$ denotes the zero-temperature Fermi occupation factor, ensuring that only particle-hole excitations between occupied and unoccupied states contribute to the susceptibility.

Interacting models, however, pose a substantially greater analytical challenge. The presence of many-body correlations generally precludes closed-form solutions, requiring the susceptibility to be computed either within controlled approximations—such as the random phase approximation (RPA) or dynamical mean-field theory (DMFT)—or by means of numerically exact techniques, including quantum Monte Carlo (QMC), density matrix renormalization group (DMRG), and exact diagonalization (ED). Among the canonical lattice models used to investigate response functions in correlated electron systems, the Hubbard model is particularly prominent: despite its deceptively simple form, it exhibits a remarkably rich spectrum of magnetic phenomena, ranging from antiferromagnetic correlations to frustrated and spin-liquid–like regimes.

In the following, we discuss in two subsections the dynamical susceptibilities of the Hubbard and SSH–Hubbard models separately, thereby illustrating how local Coulomb interactions renormalize the magnetic response in correlated systems. For the Hubbard model, we summarize well-established results from the literature and present representative susceptibility plots for completeness and comparison. The analysis of the SSH–Hubbard model, in contrast, is carried out in this work.

\subsection{Hubbard model}\label{sec:notsolvable}

The Hamiltonian of the one-dimensional (1D) Hubbard model in momentum space is given by:
\begin{equation}
H = \sum_{k, \sigma} \xi_k \, c^\dagger_{k,\sigma} c_{k,\sigma} + \frac{U}{N} \sum_{k, k^\prime, q} c^\dagger_{k-q,\uparrow} c^\dagger_{k^\prime + q,\downarrow} c_{k^\prime,\downarrow} c_{k,\uparrow},
\end{equation}
where \(\xi_k=\varepsilon_k-\mu \) is the dispersion relation (with $\varepsilon_k = -2 t \cos k$, for hopping amplitude $t$), and \( U \) represents the on-site Coulomb repulsion. Here, $N$ denotes the number of lattice sites. Despite its apparent simplicity, the Hubbard model is not exactly solvable in arbitrary dimensions, and its phase diagram remains the subject of intensive numerical and analytical investigation. 

In the following, we analyze the imaginary part of the dynamical spin susceptibility across three distinct interaction regimes. The imaginary component characterizes the spectrum of magnetic excitations and describes the dissipative response of the system, thereby providing direct information about the available spin-fluctuation modes at given momentum and frequency.

\textbf{I. Weak coupling limit ($U\ll t$):}  
In the weak-coupling regime, the magnetic response can be treated within the RPA. In this approximation, the imaginary part of the dynamical spin susceptibility is given by \cite{mahan2013many}
\begin{eqnarray}\label{eq:sushubbardweak}
\i\chi^{\mathrm{RPA}}(q,\omega)=\frac{\i\chi^{0}}{[1-U \r\chi^{0}]^{2} + [U \i\chi^{0}]^{2}},
\end{eqnarray}
where $\r\chi^{0}$ and $\i\chi^{0}$ denote the real and imaginary parts of the noninteracting susceptibility $\chi^{0}$ [see Eq.~\eqref{eq:nonint-sus}].

\begin{figure}[h]
	{\centering
		\includegraphics[scale=0.4]{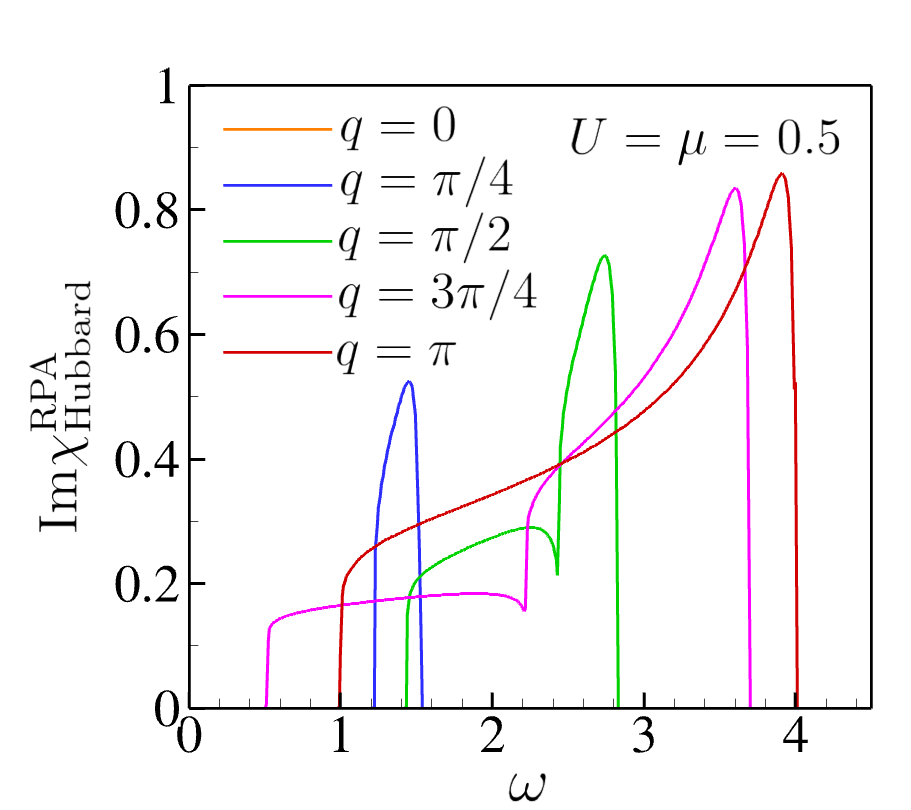}
		\includegraphics[scale=0.4]{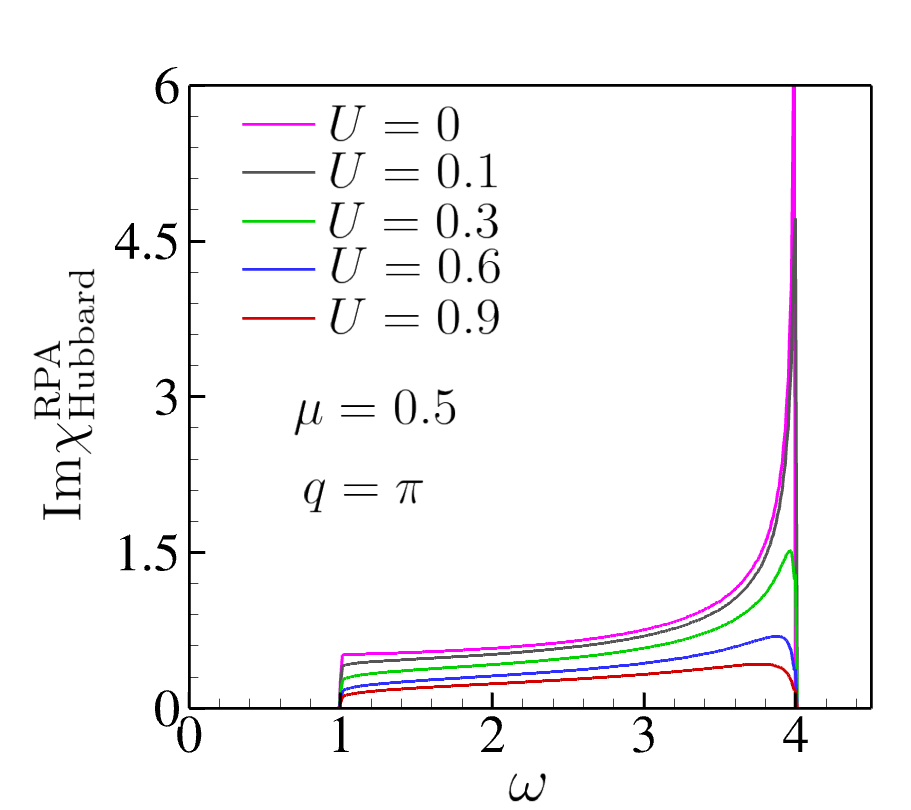}
		\caption{\small 
			Imaginary part of the dynamical spin susceptibility of the Hubbard model (energies are given in units of $t$). 
Top: $\i \chi^{\mathrm{RPA}}(q,\omega)$ for $U/t=0.5$ and $\mu/t=0.5$, corresponding to half filling ($\mu=U/2$), shown for several values of $q$. 
Bottom: $\i \chi^{\mathrm{RPA}}(q=\pi,\omega)$ for $\mu/t=0.5$ and different interaction strengths $U/t$.} \label{fig:hubbardsuspic}}
\end{figure}

The imaginary part of the susceptibility characterizes the spectrum of spin excitations and represents the dissipative response of the system. A finite value of $\i\chi^{\mathrm{RPA}}$ therefore signals the presence of particle-hole excitations at a given momentum $q$ and frequency $\omega$. In the top panel of Fig.~\ref{fig:hubbardsuspic}, we plot $\i\chi^{\mathrm{RPA}}(q,\omega)$ for several values of $q$. In this regime the interaction primarily renormalizes the excitation spectrum without generating qualitatively new modes.

For $q=0$, the particle-hole excitation energy $\varepsilon_{k+q}-\varepsilon_k$ vanishes identically since $\varepsilon_{k+q}=\varepsilon_k$. Consequently, a spatially uniform perturbation couples weakly to dynamical spin excitations, and the dissipative response is strongly suppressed. As $q$ increases from zero, the external perturbation acquires a spatial modulation, allowing finite-energy particle-hole processes and leading to a nonzero response. Increasing $q$ corresponds to probing progressively shorter wavelengths of the spin correlations. In particular, at $q=\pi$ (in units of the inverse lattice constant) the modulation wavelength becomes comparable to the lattice spacing. The shift and enhanced intensity of spectral features at larger $q$ indicate that short-wavelength excitations become increasingly important, reflecting the wavevector dependence of the underlying spin correlations.

To examine the effect of interactions, the bottom panel of Fig.~\ref{fig:hubbardsuspic} shows $\i\chi^{\mathrm{RPA}}(q=\pi,\omega)$ for several values of the interaction strength $U$. As $U$ increases, the peak amplitude decreases markedly, indicating a suppression of the spectral weight of spin excitations. In contrast, the overall frequency range over which the response occurs remains essentially unchanged. This behavior is consistent with the exact solution of the one-dimensional Hubbard model, where the spin excitation continuum at $q=\pi$ retains a bandwidth of $4t$ for all $U>0$, even though a charge gap opens in the spectrum. Thus, interactions mainly redistribute and suppress spectral weight without altering the kinematic limits determined by the noninteracting band structure.

The suppression of $\i\chi^{\mathrm{RPA}}$ with increasing $U$ can be understood physically as a consequence of the reduced mobility of electrons in the presence of on-site repulsion. Stronger interactions increase the energy cost of virtual particle-hole fluctuations, thereby reducing the phase space available for low-energy dynamical excitations. As a result, the dissipative magnetic response is progressively suppressed with increasing interaction strength, except near the RPA instability condition $1-U \r\chi^0(q,\omega)=0$.

\textbf{II. Intermediate-coupling regime (\( U \sim t \)):} 
In the intermediate-coupling regime, simple analytical approximations such as RPA are no longer sufficient to describe the dynamical response accurately, since interaction effects substantially renormalize the excitation spectrum. In this regime, short-range many-body correlations become important, and numerical approaches are required. Using cluster perturbation theory, Raum \textit{et al.}~\cite{raum2020two} computed the imaginary part of the dynamical spin susceptibility of the one-dimensional Hubbard model beyond the weak-coupling limit.

Their results (see Fig.~4 of Ref.~\onlinecite{raum2020two}) show that for $U=1$ and $U=2$ (in units of $t$), the susceptibility at $q=\pi$ exhibits a pronounced maximum with a peak value of approximately $0.5$. Relative to the weak-coupling regime, the spectral features are noticeably renormalized, indicating that interaction effects already play a significant role at intermediate coupling.

Moreover, as $q$ increases toward $\pi$, the peak in $\i\chi(q,\omega)$ becomes broader and its amplitude increases. This trend indicates that spin fluctuations are enhanced at larger momenta, consistent with the growing importance of short-wavelength antiferromagnetic correlations. The broadening of the response also shows that the excitations are no longer well described by a simple weakly interacting particle-hole picture, but instead reflect the increasing influence of many-body correlations.

Overall, the cluster perturbation theory results demonstrate that the intermediate-coupling regime is characterized by a substantial redistribution of spectral weight and a momentum-dependent enhancement of the spin response, especially near $q=\pi$, where antiferromagnetic correlations are strongest.

\begin{figure}[h]
{\centering
\includegraphics[scale=0.4]{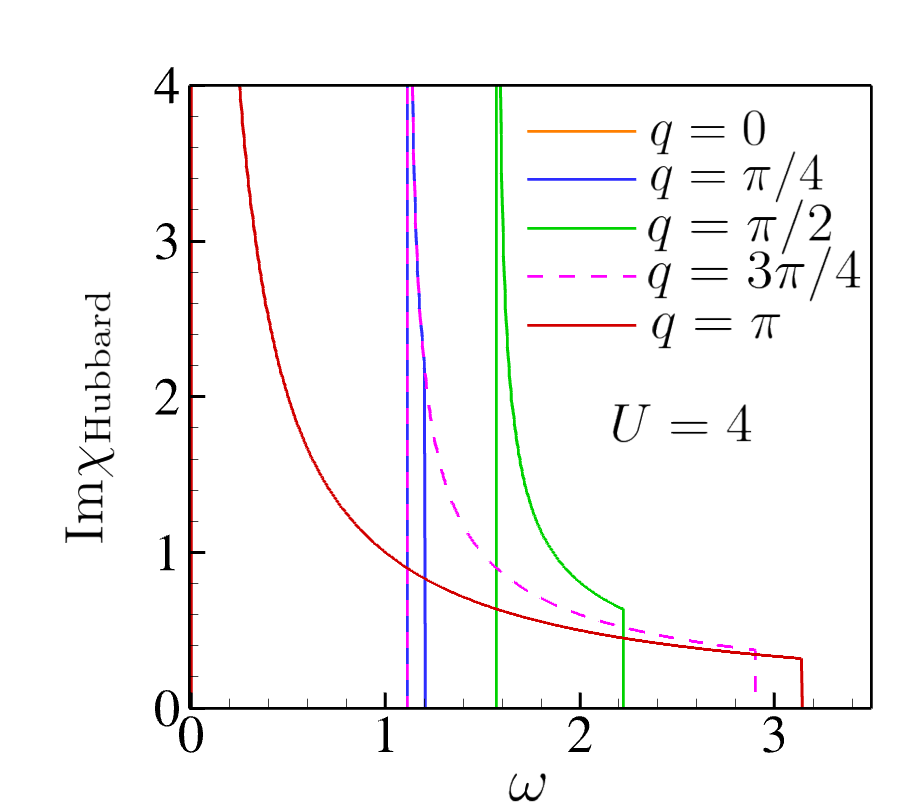}
\includegraphics[scale=0.4]{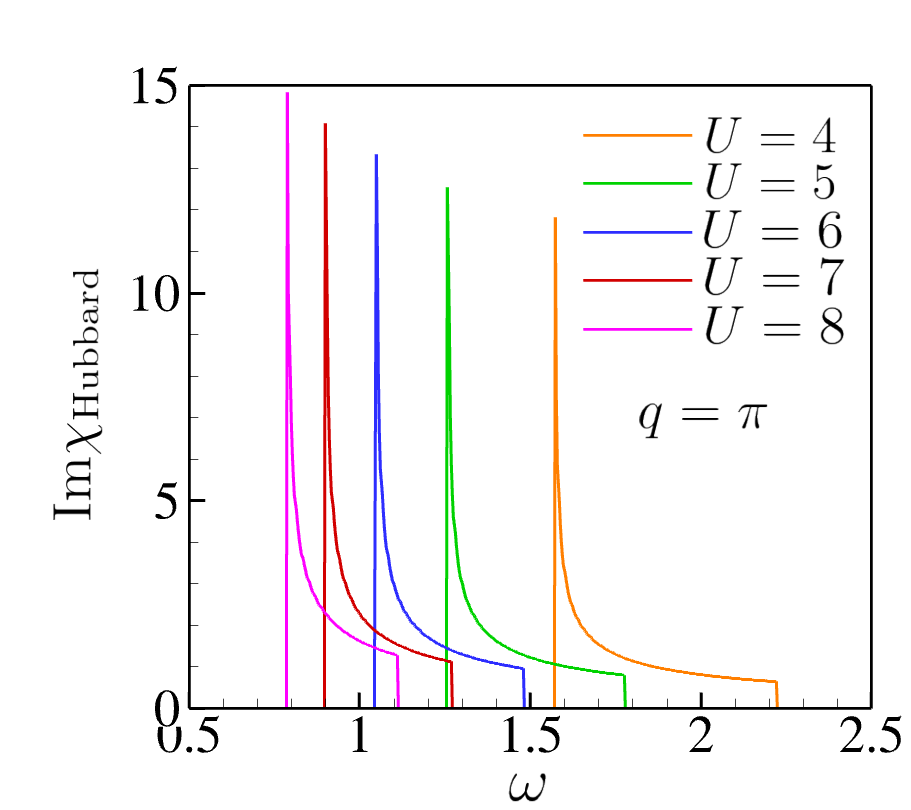}
\caption{\small
Imaginary part of the dynamical spin susceptibility of the Hubbard model in the strong-coupling limit (energies are given in units of $t$). 
Top: $\i\chi(q,\omega)$ for $U/t=4$ (corresponding to $J=t$) for several values of $q$. 
Bottom: $\i\chi(q=\pi,\omega)$ for different interaction strengths $U/t$.
} \label{fig:hubbardsuspicstrong}}
\end{figure}	

\textbf{III. Strong-coupling limit ($U\gg t$):}  
In the strong-coupling limit ($U\gg t$) at half filling ($\mu=U/2$), charge fluctuations are strongly suppressed and the one-dimensional Hubbard model maps onto an antiferromagnetic spin-$1/2$ Heisenberg chain with nearest-neighbor superexchange interaction $J=4t^2/U$. In this regime the low-energy dynamics are governed predominantly by spin degrees of freedom. The dynamical spin response of the Heisenberg chain is well captured by the Müller ansatz, which provides an approximate expression for the dynamical spin structure factor $S(q,\omega)$ \cite{raum2020two},
\begin{equation}
S(q,\omega)\approx
\frac{
\Theta[\omega-\omega_L(q)]
\Theta[\omega_U(q)-\omega]
}{
\sqrt{\omega^2-\omega_L^2(q)}
}.
\end{equation}
Here, $\Theta$ denotes the Heaviside function and the lower and upper boundaries of the two-spinon continuum are
\begin{equation}
\omega_L(q)=\frac{\pi J}{2}|\sin q|,
\qquad
\omega_U(q)=\pi J\left|\sin\frac{q}{2}\right|.
\end{equation}
At zero temperature the imaginary part of the dynamical spin susceptibility is directly related to the structure factor through
\begin{equation}
\i\chi(q,\omega)=\pi S(q,\omega),
\end{equation}
so that the Müller ansatz provides an approximate description of the dissipative part of the dynamical susceptibility within the two-spinon continuum bounded by $\omega_L(q)$ and $\omega_U(q)$. Although approximate, it reproduces the overall spectral structure and the characteristic singular behavior near the lower boundary of the continuum, while quantitatively precise results require more advanced numerical approaches such as DMRG calculations \cite{dmrg,nocera2018finite}. The imaginary part of the susceptibility is nonzero only within the frequency interval $[\omega_L(q),\omega_U(q)]$, which defines the two-spinon excitation continuum. At $q=0$, both boundaries vanish identically, $\omega_L=\omega_U=0$, implying the absence of finite-frequency spectral weight, $\i\chi(q=0,\omega)=0$. This reflects the fact that a spatially uniform perturbation cannot efficiently excite finite-energy spin fluctuations in the Heisenberg limit.

The top panel of Fig.~\ref{fig:hubbardsuspicstrong} shows $\i\chi(q,\omega)$ for several values of $q$ at $U/t=4$, corresponding to $J=t$. As $q$ increases, the spectral weight shifts to higher frequencies and the width of the excitation continuum increases. Physically, increasing momentum transfer probes progressively shorter-wavelength spin correlations. Near $q=\pi$, antiferromagnetic correlations become dominant, leading to a strong enhancement of the response. At this wavevector,
$\omega_L(\pi)=0$, and $\omega_U(\pi)=\pi J$,
and the susceptibility develops the characteristic low-frequency singularity $\i\chi(q=\pi,\omega)\sim \frac{1}{\omega}$,
which is a hallmark of the spinon continuum in one dimension.

Unlike higher-dimensional antiferromagnets, the one-dimensional spin-$1/2$ Heisenberg chain does not possess long-range magnetic order in its ground state due to strong quantum fluctuations. Its elementary excitations are fractionalized spinons carrying spin $1/2$, rather than conventional magnon excitations with spin $1$. Consequently, the dynamical spin response forms a broad continuum rather than a sharp dispersing mode. The Müller ansatz successfully captures the essential features of this continuum, including the continuum boundaries and the singular spectral enhancement near $q=\pi$.

In the bottom panel of Fig.~\ref{fig:hubbardsuspicstrong}, we plot $\i\chi(q=\pi,\omega)$ for several interaction strengths $U/t$. As $U$ increases, the effective exchange interaction $J=4t^2/U$ decreases, causing the entire spin excitation spectrum to shift toward lower frequencies. Consequently, the frequency range over which the susceptibility is finite becomes progressively narrower. At the same time, the peak amplitude increases due to the compression of spectral weight into a smaller energy window. This behavior reflects the reduction of the characteristic magnetic energy scale in the strong-coupling regime. 

This strong-coupling behavior differs qualitatively from the weak-coupling RPA regime, where the spectral bandwidth remains primarily controlled by the electronic hopping scale $t$. In contrast, in the Heisenberg limit the spin dynamics are governed entirely by the superexchange scale $J\propto t^2/U$, leading to a systematic narrowing of the spin excitation continuum as $U$ increases.

\subsection{SSH--Hubbard model}\label{sec:notsolvablesshh}

Recent studies of the SSH--Hubbard model have provided important insights into the interplay between topology and interactions. 
Using edge degeneracy and edge entanglement entropy, it was shown that the system realizes a topological phase for $v<w$ and a trivial phase for $v>w$, with the Hubbard interaction strength varied in the range $0\le U\le 2$~\cite{ye2016entanglement,wang2015detecting}.
Consistent results were obtained by computing bulk topological invariants from Green’s functions extracted via DMRG, where the invariant takes the value $2$ in the topological phase and $0$ in the trivial phase~\cite{manmana2012topological}.

Beyond purely electronic correlations, the interplay between Hubbard interaction and lattice degrees of freedom has also been extensively explored.
In particular, when the Hubbard interaction is combined with electron-phonon coupling on a square lattice, a first-order transition between a Neel phase and a bond-ordered-wave phase was observed as the electron-phonon coupling strength increases.
At weak electron-phonon coupling, the system reproduces the behavior of the Hubbard model without lattice coupling~\cite{feng2022phase,xing2021quantum,cai2021antiferromagnetism,gotz2022valence,sous2018light,wang2025robust,yang2022functional}.

The role of lattice coupling becomes even richer in the presence of attractive interactions.
The interplay between attractive Hubbard interactions and lattice degrees of freedom has been extensively investigated, as pairing tendencies compete directly with charge and bond ordering.
In particular, the SSH model with an attractive Hubbard interaction exhibits a rich phase diagram featuring charge-density-wave order, bond-ordered phases, and s-wave superconducting pairing.
Determinant quantum Monte Carlo studies have shown that charge order dominates at weak electron-phonon coupling, while bond order becomes favorable at strong coupling. 
Upon doping away from half-filling, both ordered phases are suppressed in favor of dominant s-wave pairing correlations. Importantly, the SSH-type electron-phonon interaction competes with the attractive Hubbard interaction and significantly suppresses the strength of s-wave pairing correlations~\cite{xing2023attractive}.

To characterize the competing ordered phases discussed above, response functions provide a natural and physically transparent diagnostic. For example, the static susceptibility of the two-band Hubbard model—widely employed as an effective description of the correlated semiconductor FeSi—has proven valuable for probing correlation effects ~\cite{urasaki1998thermal}. Here, we examine the dynamical susceptibility of the SSH--Hubbard model by evaluating the non-interacting Lindhard bubble and including interaction effects within the RPA.

The one-dimensional SSH--Hubbard model is described by the Hamiltonian
\begin{equation}
H = H_{\mathrm{SSH}} + U \sum_{j} \left(n^a_{j \uparrow} n^a_{j \downarrow} + n^b_{j \uparrow} n^b_{j \downarrow}\right),
\end{equation}
where $n^\alpha_{j\sigma}=\alpha^{\dagger}_{j\sigma}\alpha_{j\sigma}$ with $\alpha=a,b$ labeling the two sublattices of the unit cell, and $U$ denotes the on-site Coulomb interaction. The term $H_{\mathrm{SSH}}$ represents the noninteracting SSH Hamiltonian defined in Eq.~\eqref{s1.1}.

For the noninteracting SSH model the Bloch spectrum consists of two bands with dispersion
\begin{equation}\label{eq:sshenergy}
\epsilon^{\pm}_k=\pm \sqrt{v^2+w^2+2vw\cos k}.
\end{equation}
The corresponding Bloch eigenstates can be written as
\begin{align}\label{Eq:1-particle}
\ket{u_{k}^{\pm}} =\frac{1}{\sqrt{2}}
\begin{pmatrix}
e^{-i\phi_k /2}\\[4pt]
\pm e^{i\phi_k /2}
\end{pmatrix},
\end{align}
where the phase $\phi_k$ is defined through
\begin{equation}
e^{i\phi_k}=\frac{v+w e^{-ik}}{|v+w e^{-ik}|}.
\end{equation}

In the weak-coupling regime, where the interaction strength is small compared to the hopping amplitudes ($U/v \ll 1$ and $U/w \ll 1$), the dynamical response of the system can be treated within the RPA. The key quantity entering this approach is the noninteracting susceptibility of the SSH bands, which can be written as
\begin{eqnarray}\label{eq:sus-SSH--Hubbard}
\chi^{0}_{\mathrm{SSH}}(q, \omega) =
\frac{2}{N} \sum_{k,\nu,\nu'} 
\frac{ \Theta(-\epsilon_k^{\nu}) - \Theta(-\epsilon_{k+q}^{\nu'})}{\omega+i\eta+ \epsilon_{k}^{\nu} - \epsilon_{k+q}^{\nu'} }
F^{\nu,\nu'}_{k,k+q},~~~~
\end{eqnarray}
where $\nu,\nu'=\pm$ label the two SSH bands and the factor of $2$ accounts for spin degeneracy. The step functions $\Theta(-\epsilon_k^\nu)$ represent the zero-temperature occupation of the single-particle states. At half filling the chemical potential is fixed at $\mu=0$, so that states with $\epsilon_k^\nu<0$ are occupied while those with $\epsilon_k^\nu>0$ are empty. The quantity
$
F^{\nu,\nu'}_{k,k+q}=|\langle u_{k}^{\nu} | u_{k+q}^{\nu'}\rangle|^2
$
is the band form factor describing the overlap between Bloch states in bands $\nu$ and $\nu'$.

Substituting the eigenstates in Eq.~(\ref{Eq:1-particle}) into the definition of the form factor yields
\begin{equation}
F^{\nu,\nu'}_{k,k+q}=\frac{1}{2}\left[1+\nu\nu' \cos(\phi_k-\phi_{k+q})\right],
\end{equation}
which encodes the sublattice pseudospin structure of the SSH Bloch states and governs the relative weight of intra- and interband particle-hole excitations.

\begin{figure}[h]
	{\centering
		\includegraphics[scale=0.4]{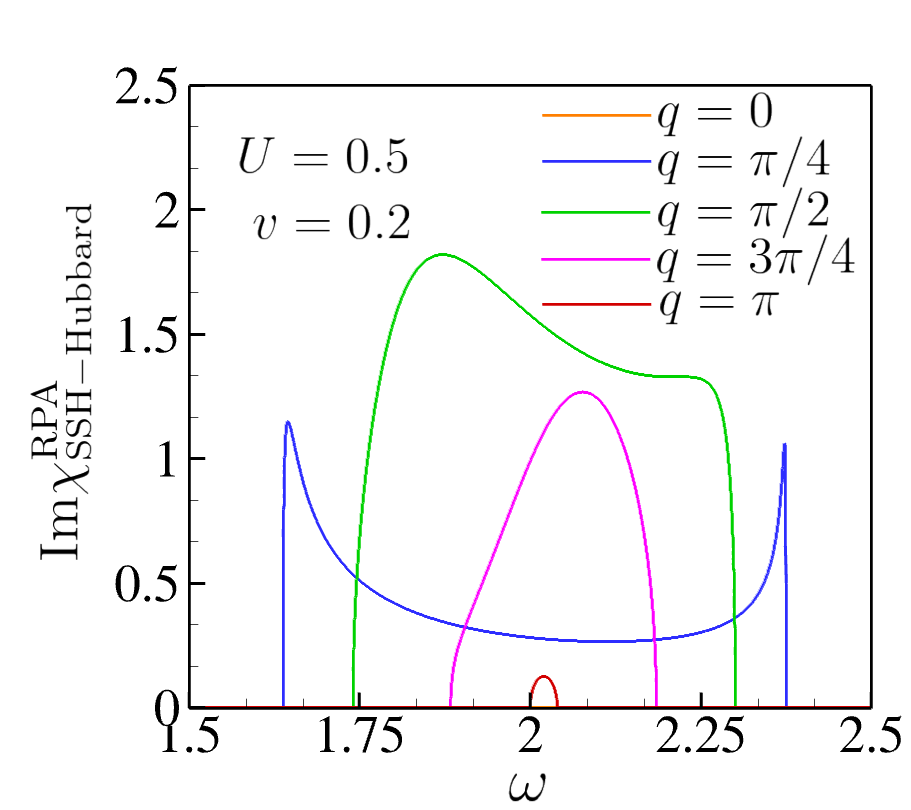}
		\includegraphics[scale=0.4]{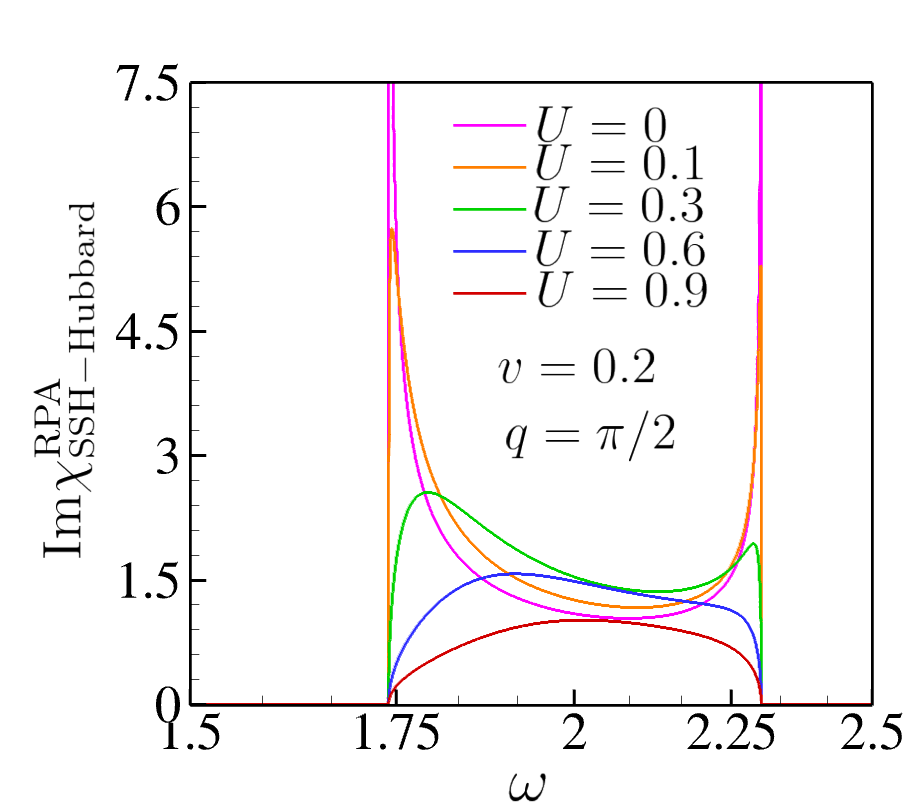}
		\caption{\small 
			Imaginary part of the dynamical susceptibility for the SSH--Hubbard model.
Top panel: $\i\chi(q,\omega)$ for fixed interaction strength
$U/w=0.5$ (with $w=1$) and several wave vectors $q$, using
$v/w=0.2$. Bottom panel: $\i\chi(q,\omega)$ at fixed wave vector
$q=\pi/2$ for different values of the Hubbard interaction $U$.}
\label{fig:sussshh} }
\end{figure}
As shown in Fig.~\ref{fig:sussshh} (bottom panel), the imaginary part of the
dynamical susceptibility of the noninteracting SSH model (pink curve) at
$q=\pi/2$ exhibits two pronounced peaks near the lower and upper boundaries
of the particle-hole continuum. This behavior contrasts with the
half-filled free-fermion chain at $q=\pi$, where $\i\chi$
displays a single dominant peak near the upper continuum edge. The
relevant wave vectors reflect the characteristic scales of the two
systems. For the half-filled free-fermion chain the dominant response
occurs at $q=2k_{\rm F}=\pi$, while in the SSH model the doubled unit cell
reduces the Brillouin zone to $[-\pi/2,\pi/2]$, giving the corresponding
characteristic wave vector $q=2k_{\rm F}=\pi/2$. The two-peak structure of
the SSH response originates from its two-band spectrum, which produces
distinct extrema in the interband particle-hole excitation continuum.

When the Hubbard interaction is included, these noninteracting spectral
features become substantially suppressed. Figure~\ref{fig:sussshh} (top
panel) shows $\i\chi^{\mathrm{RPA}}(q,\omega)$ obtained from
Eq.~\eqref{eq:sushubbardweak}. At $q=0$, the density operator does not
change the crystal momentum of the excitation, which strongly restricts
the available finite-energy particle-hole processes and leads to a
pronounced suppression of $\i\chi^{\mathrm{RPA}}(0,\omega)$.

Throughout the SSH--Hubbard calculations we set $\mu=0$, corresponding to
half filling. In the noninteracting limit, the dominant particle-hole
processes occur near the characteristic wave vector
$q=2k_{\rm F}=\pi/2$. At this wave vector, occupied states in the lower
band can be connected efficiently to unoccupied states in the upper band,
maximizing the available phase space for excitations. Consequently, the
susceptibility develops enhanced and broadened spectral weight around
$q=\pi/2$, while the response remains comparatively weak at $q=0$ and near
the Brillouin-zone boundary.

Figure~\ref{fig:sussshh} further shows that
$\i\chi^{\mathrm{RPA}}(q,\omega)$ decreases systematically with
increasing interaction strength $U$. Both the peak intensity and the total
spectral weight are reduced, indicating suppression of charge fluctuations
by the on-site repulsion. Within the RPA framework this suppression arises
through the interaction-dependent denominator
$\chi^{\mathrm{RPA}}(q,\omega)=\frac{\chi^0(q,\omega)}{1-U\chi^0(q,\omega)}$,
which renormalizes the collective response while leaving the boundaries of
the underlying particle-hole continuum largely unchanged. Consequently,
the overall frequency range in which the spectral function is nonzero
remains approximately fixed, even though the spectral intensity is
strongly reduced. This behavior closely parallels that of the uniform
Hubbard chain shown in Fig.~\ref{fig:hubbardsuspic}, indicating that in
the weak-coupling regime the interaction primarily redistributes spectral
weight rather than modifying the continuum support.

For intermediate and strong interactions ($U\sim w$ and $U\gg w$, where
$w$ denotes the intercell hopping amplitude), the RPA description becomes
inadequate and nonperturbative approaches are required. Exact-diagonalization
studies have shown that the topological phase of the SSH--Hubbard model
remains stable against moderate interactions and can be identified through
degeneracies in the entanglement spectrum~\cite{ye2016entanglement}.
Lanczos calculations over a broad interaction range further reveal the
opening of a correlation-induced charge gap and a redistribution of
spectral weight, with the excitation spectrum evolving toward
Hubbard-like bands with renormalized effective parameters~\cite{mikhail2024schrieffer}.

In the strong-coupling regime, charge fluctuations become strongly
suppressed and the low-energy response is dominated by correlated spin
dynamics on an effectively dimerized lattice. The resulting spectral
features differ qualitatively from both the weakly interacting SSH model
and the uniform Hubbard chain: high-energy charge excitations lose
spectral weight, while the remaining low-energy structures are governed
by interaction-renormalized hopping processes and short-range spin
correlations. Reliable evaluation of the dynamical susceptibility in this
regime therefore requires fully correlated numerical methods such as
exact diagonalization or DMRG.

\section{Hatsugai--Kohmoto (HK) and extended SSH--HK models}

\subsection{HK model}\label{sec:solvable}
The HK model describes strongly correlated electrons with an infinite-range
interaction. The Hamiltonian reads
\begin{eqnarray}
\label{eq:hkreal}
H &=& -t \sum_{\langle i,j\rangle,\sigma}
\left(c^{\dagger}_{i\sigma}c_{j\sigma}+\mathrm{H.c.}\right) \nonumber\\
&&+ \frac{U}{N}\sum_{j_1,j_2,j_3,j_4}
\delta_{j_1+j_3,\,j_2+j_4}\,
c^{\dagger}_{j_1\uparrow}c_{j_2\uparrow}
c^{\dagger}_{j_3\downarrow}c_{j_4\downarrow}.
\end{eqnarray}
The first term describes nearest-neighbor hopping with amplitude $t$, and
$N$ is the number of lattice sites. The Kronecker delta
$\delta_{j_1+j_3,\,j_2+j_4}$ restricts the summation to lattice indices
satisfying $r_{j_1}+r_{j_3}=r_{j_2}+r_{j_4}$, i.e., it preserves the
center-of-mass coordinate of an opposite-spin pair and renders the
interaction effectively infinite ranged \cite{hatsugai1992exactly}.

The interaction term can be viewed as a correlated two-particle scattering
process in which two fermions are annihilated and two are created while
conserving the total lattice momentum. Equivalently, the interaction
redistributes occupation among momentum states without changing the
pair center-of-mass motion. Although the HK Hamiltonian appears nonlocal in
real space when written in terms of fermionic operators, it admits an
equivalent local formulation in terms of real-space current operators,
consistent with its $Z_2$ symmetry-breaking universality class
\cite{bai2026local}. In this formulation the Hamiltonian is expressed
through local currents constructed from fermionic bilinears. In one
dimension such currents obey a Kac--Moody (affine) algebra; in the present
case they satisfy an $su(2)_1$ current algebra. Using the
Bjorken--Johnson--Low (BJL) prescription, which relates the high-frequency
behavior of time-ordered correlators to equal-time commutators, the
anomalous commutation relations of these currents can be derived
explicitly. Within this framework it has been shown that physical
observables, such as the charge susceptibility, coincide with those
obtained in the original fermionic formulation, establishing an equivalent
local real-space representation of the HK model \cite{bai2026local}.

The HK model exhibits an interaction-driven metal-insulator transition.
At half filling the transition occurs when the interaction reaches the
noninteracting bandwidth, $U_c=4t$: the system is metallic for $U<U_c$ and
insulating for $U>U_c$. Away from half filling, it remains metallic for all
interaction strengths.

In momentum space, the grand-canonical Hamiltonian
$K_{\mathrm{HK}}=H-\mu N$ decomposes into independent momentum sectors,
\begin{eqnarray}
\nonumber K_{\mathrm{HK}} &=& \sum_k K_{\mathrm{HK}}(k),\\
K_{\mathrm{HK}}(k)&=&
\xi_k (n_{k\uparrow}+n_{k\downarrow}) + U\, n_{k\uparrow}n_{k\downarrow},
\label{Eq:HK-Hamilton-Kspace}
\end{eqnarray}
where $\xi_k=\varepsilon_k-\mu$ with $\varepsilon_k=-2t\cos k$ and
$n_{k\sigma}=c^\dagger_{k\sigma}c_{k\sigma}$. The interaction therefore
couples only fermions with the same momentum $k$, making each $k$ sector
exactly solvable.

Because different momentum sectors decouple, the HK model belongs to the small
class of interacting systems for which the exact single-particle Green's
function can be obtained analytically. As a result, vertex corrections vanish,
and Wick-type factorization remains valid in evaluating linear-response
quantities. Consequently, charge and spin susceptibilities can be computed
directly from the exact interacting Green's function in the same formal manner
as in the noninteracting limit
\cite{phillips2018absence,nogueira1996study}.
The exact Green's function for the HK model is
\begin{equation}
\label{eq:hkgreens}
G_{\mathrm{HK}}(k,\omega)=\frac{1-g_k(U)}{\omega-\xi_k}+
\frac{g_k(U)}{\omega-(\xi_k+U)},
\end{equation}
where the zero-temperature occupation function $g_k(U)$ is defined through
the Heaviside step function $\Theta(x)$ as
\begin{equation}
\label{eq:g(k,U)}
g_k(U)=\Theta(-\xi_k)\left[\Theta(|\xi_k|-U)
+\frac{1}{2}\Theta(U-|\xi_k|)
\right].
\end{equation}

Equation~(\ref{eq:g(k,U)}) implies three distinct regimes. For
$\xi_k<-U$, one has $g_k(U)=1$, corresponding to a fully occupied state.
For $-U<\xi_k<0$, the occupation becomes fractional with $g_k(U)=1/2$.
Finally, for $\xi_k>0$ the state is empty and $g_k(U)=0$. These
discontinuities at $\xi_k=0$ and $|\xi_k|=U$ reflect the characteristic
spectral structure of the HK model and underlie the unconventional
metal-insulator transition discussed below.

Using the exact Green’s function [Eq.~\eqref{eq:hkgreens}], the dynamical spin
susceptibility can be written in the compact branch representation
\begin{equation}
\label{eq:sushkcompact}
\chi_{\mathrm{HK}}(q,\omega)=\frac{1}{N}\sum_{k,\sigma}\sum_{ss'=0,1}W^{ss'}_{k,k+q}
\frac{\Theta\big[-E^{(s')}_{k+q}\big]-\Theta\big[-E^{(s)}_{k}\big]}{\omega+i\eta+E^{(s)}_k-E^{(s')}_{k+q}},
\end{equation}
where $E_k^{(s)}=\xi_k+sU$. The transition weights are
\(
W^{ss'}_{k,k+q}=w_k^{(s)}w_{k+q}^{(s')}
\),
constructed from the spectral weights $w_k^{(s)}$ that distribute the
Green's function residue between the two HK excitation branches:
$w_k^{(0)}=1-g_k(U)$ and $w_k^{(1)}=g_k(U)$, with
$w_k^{(0)}+w_k^{(1)}=1$. Equation~\eqref{eq:sushkcompact} expresses
$\chi_{\mathrm{HK}}(q,\omega)$ as a sum of $k$‑resolved transitions between the
two HK spectral branches \cite{phillips2018absence}.

\begin{figure}[h]
	{\centering
		\includegraphics[scale=0.4]{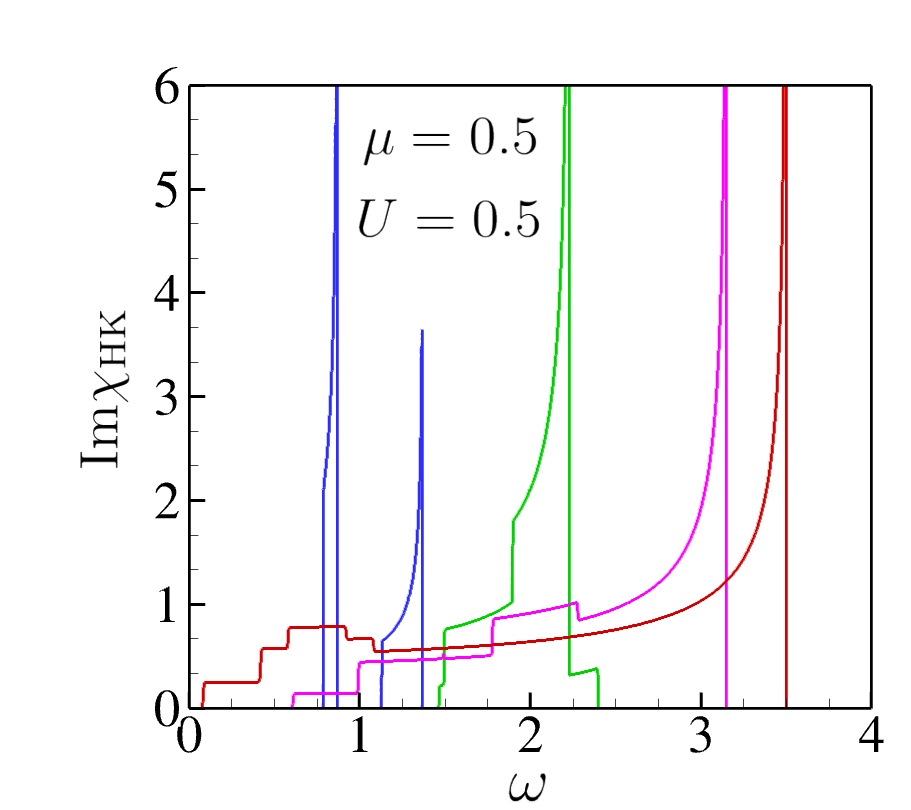}
		\includegraphics[scale=0.4]{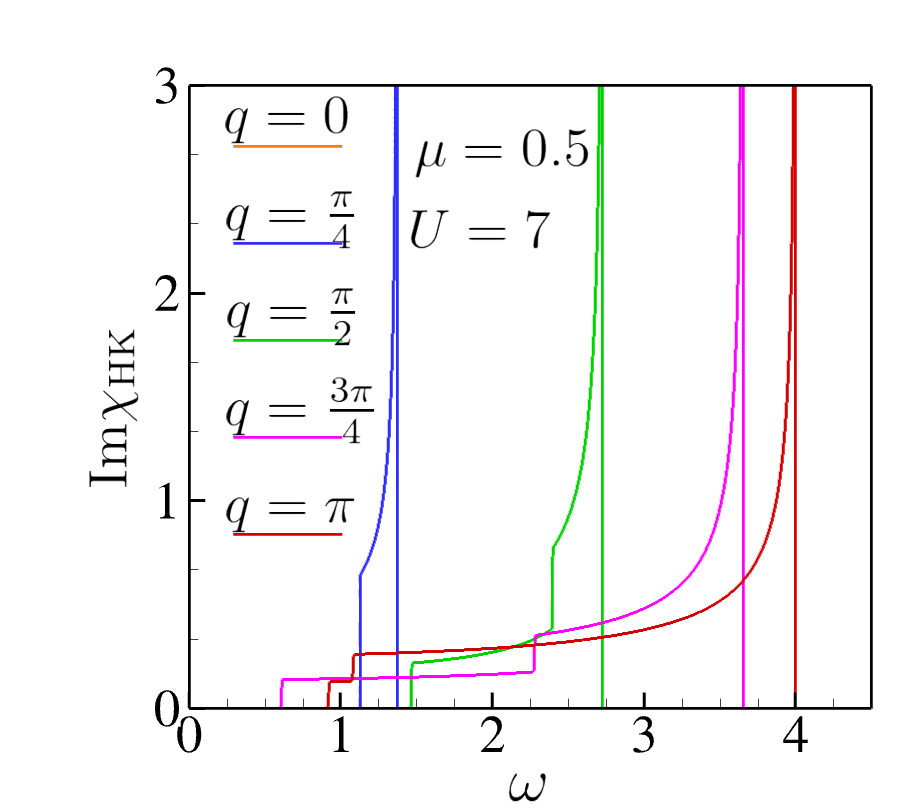}
		\caption{Imaginary part of the dynamical spin susceptibility of the HK model for several wave vectors $q$. The chemical potential is set to $\mu/t=0.5$. Top: weak interaction, $U/t=0.5$. Bottom: strong interaction, $U/t=7$.	} 
\label{fig:hksusfig}}
\end{figure}

Fig.~\ref{fig:hksusfig} presents the imaginary part of the HK model
susceptibility as a function of frequency $\omega$ for several wavevectors
$q$. The top panel corresponds to $U/t=0.5$, while the bottom panel shows
results for a stronger interaction, $U/t=7$. A key distinction between the
exactly solvable HK model and the Hubbard and SSH--Hubbard models is that
the dynamical susceptibility of the HK model is known exactly even in the
strong-coupling regime and is given by Eq.~\eqref{eq:sushkcompact}. This
follows from the momentum-space locality of the interaction: different
$k$ sectors remain decoupled, vertex corrections vanish, and the response
can therefore be expressed exactly in terms of the interacting HK Green's
function.

For $q=0$, the denominators of the first and fourth terms in
Eq.~\eqref{eq:sushkcompact}, together with the numerators of the second and
third terms, vanish identically, which leads to
$\i\chi_{\rm HK}(q=0,\omega)=0$. Physically, this reflects that a
spatially uniform perturbation cannot induce finite-energy particle-hole
excitations.

As $q$ increases, the frequency interval over which the susceptibility is
nonzero broadens, indicating that fermionic excitations become accessible
over an extended energy window. In the HK model the interaction is local in
momentum space and therefore does not mix different $k$ states. The
susceptibility can consequently be expressed as a sum of independent
excitation processes connecting occupied and unoccupied states with
momenta $k$ and $k+q$. Although the HK model is not a conventional Fermi
liquid, the dynamical response retains the structure of transitions
between the two spectral branches at energies $\xi_k$ and $\xi_k+U$. As
$q$ increases the wavelength of the corresponding spatial modulation
decreases. At the Brillouin-zone boundary $q=\pi$ the perturbation becomes
staggered, maximizing its coupling to short-wavelength fluctuations.

In contrast to the Hubbard model (Fig.~\ref{fig:hubbardsuspic}, top panel),
the HK model exhibits a significantly larger $\i\chi_{\rm HK}(q,\omega)$.
This enhancement originates from the structure of the HK Green's function,
which contains two sharp poles corresponding to excitations at energies
$\xi_k$ and $\xi_k+U$ [Eq.~\eqref{eq:hkgreens}]. Consequently, the susceptibility is composed of multiple inter-branch excitation processes arising from transitions between these two spectral branches. In contrast, the weak-coupling Hubbard Green’s function, as defined in Eq.~\eqref{eq:sushubbardweak}, features a single quasiparticle pole; this restriction limits the available transitions and leads to a significantly smoother evolution of the spectral weight.

The interaction-driven metal-insulator transition at half filling is
directly reflected in the behavior of the dynamical susceptibility. In the
metallic phase ($U<4t$), the two spectral branches overlap in energy,
facilitating low-energy excitations and maintaining finite spectral weight
in $\i\chi_{\rm HK}(q,\omega)$ down to small frequencies. As $U$
increases beyond the critical value $U_c=4t$, these branches separate
completely and an interaction gap opens between the occupied and
unoccupied states. In this insulating regime, transitions between the
branches require a finite minimum energy, leading to a characteristic
threshold behavior where $\i\chi_{\rm HK}(q,\omega)$ vanishes below a
frequency corresponding to the interaction-induced gap.

The step-like structures observed in $\i\chi_{\rm HK}(q,\omega)$ at
certain values of $q$ originate from the definition of the function
$g_k(U)$, which introduces sharp boundaries in momentum space through
Heaviside functions. As the frequency increases, new regions of phase
space satisfying the excitation conditions become accessible. 
Whenever such a region is accessed, additional inter-branch transitions abruptly contribute to the susceptibility, resulting in the characteristic step-like increases in $\mathrm{Im}\chi_{\rm HK}(q,\omega)$. Such sharp features are absent in the Hubbard model, where the spectral weight evolves smoothly due to the continuous nature of its particle-hole excitation spectrum.

Unlike in the Hubbard model, where $q$ controls the dispersion and
spectral weight of collective modes through the particle-hole continuum,
in the HK model $q$ primarily determines which excitation processes
between $k$ and $k+q$ contribute to the response. This decoupling of
momentum sectors is a direct consequence of the momentum-space locality of
the HK interaction.

In the strong-interaction regime (Fig.~\ref{fig:hksusfig}), the peak of the
susceptibility shifts to larger values of $\omega$. This behavior reflects
the interaction-induced increase of the excitation energies: as $U$
grows, the separation between the spectral branches $\xi_k$ and
$\xi_k+U$ becomes larger, so that the dominant transitions occur at
higher frequencies.

\subsection{SSH--HK model}\label{sec:solvablesshhk}

The Su--Schrieffer--Heeger model with Hatsugai--Kohmoto interaction (SSH--HK)
combines the topological band structure of the SSH chain with a
momentum-diagonal interaction, providing an exactly solvable framework for 
investigating the interplay between topology and long-range interactions 
\cite{mohamadi2025emergence}.

The total Hamiltonian is $H_{\mathrm{SSH-HK}} = \sum_k H_{\mathrm{SSH-HK}}(k)$,
consisting of the SSH term $H_{\mathrm{SSH}}(k)$ [Eq.~\eqref{s1.1}] and an 
HK-type interaction $\tilde{H}_{\mathrm{HK}}(k)$ defined in the two-sublattice
basis. Using the Fourier transforms $a_j = N^{-1/2} \sum_k e^{ikj} a_k$ and 
$b_j = N^{-1/2} \sum_k e^{ikj} b_k$, the interaction term becomes
\begin{equation}
\tilde{H}_{\mathrm{HK}}(k) =
U\left(
n^a_{k\uparrow}n^a_{k\downarrow}
+n^b_{k\uparrow}n^b_{k\downarrow}
+n^a_{k\uparrow}n^b_{k\downarrow}
+n^b_{k\uparrow}n^a_{k\downarrow}
\right),
\end{equation}
where $n^{a,b}_{k\sigma}$ are the corresponding number operators. Unlike the 
single-band model $H_{\mathrm{HK}}$ [Eq.~\eqref{Eq:HK-Hamilton-Kspace}], the 
SSH--HK interaction couples the $a$ and $b$ sublattices. 

As reported previously \cite{mohamadi2025emergence}, the ground-state phase 
diagram is significantly enriched by these interactions. At integer fillings 
$n=1$ and $n=2$, topological and trivial non-Fermi-liquid (NFL) phases 
emerge, characterized by many-body Zak phases of $2\pi$ and $0$, respectively. 
The topological NFL phase, in particular, exhibits electronic polarization 
properties analogous to those of the noninteracting SSH model 
\cite{mohamadi2025emergence}.

It is sometimes convenient to express the SSH--HK model in the band basis,
where the single-particle SSH Hamiltonian is diagonal and the HK interaction
remains momentum diagonal \cite{mai2023topological}. In this representation,
\begin{align}
&H_{\mathrm{SSH-HK}}^{\mathrm{diag}}
=\sum_{k,\sigma}\left(\epsilon_{k}^+\, n^\alpha_{k\sigma}+\epsilon_{k}^-\, n^\beta_{k\sigma}\right)
\nonumber\\
&~~+ U\sum_k\left(n^\alpha_{k\uparrow}n^\alpha_{k\downarrow}
+n^\beta_{k\uparrow}n^\beta_{k\downarrow}
+n^\alpha_{k\uparrow}n^\beta_{k\downarrow}
+n^\beta_{k\uparrow}n^\alpha_{k\downarrow}\right),
\label{eq:SSHHK-kspace_diag}
\end{align}
with many-body states written in the occupation-number basis built from the band
operators,
\begin{align}
\ket{n_1,n_2,n_3,n_4}=
(\alpha_{k\uparrow}^\dagger)^{n_1}
(\alpha_{k\downarrow}^\dagger)^{n_2}
(\beta_{k\uparrow}^\dagger)^{n_3}
(\beta_{k\downarrow}^\dagger)^{n_4}
\ket{\mathbf 0}.
\label{Eq:diagonalbases}
\end{align}
The diagonalized Hamiltonian $H_{\mathrm{SSH-HK}}^{\mathrm{diag}}$ has
the same eigenvalues as the original $H_{\mathrm{SSH-HK}}$, but the
unitary transformation to the band basis absorbs the momentum dependence of the
Bloch eigenvectors. Consequently, $H_{\mathrm{SSH-HK}}^{\mathrm{diag}}$
is suitable for quantities determined solely by the energy spectrum, whereas
observables that depend explicitly on the momentum-resolved eigenvectors—such
as the particle number in the orbital basis, the Zak phase, electric
polarization, the Luttinger integral, and the full spectral function—must be
evaluated using the original Hamiltonian $H_{\mathrm{SSH-HK}}$.

Beyond ground-state topology, the interacting SSH--HK model exhibits rich
dynamical behavior. Exact diagonalization reveals non-quasiparticle spectral
functions and density of states with clear deviations from Fermi-liquid theory
\cite{mohamadi2025emergence}.

Because the SSH--HK Hamiltonian remains diagonal in momentum even with
interaction, Wick's theorem applies, and the Matsubara susceptibility is
\begin{equation}
\chi^{\alpha\alpha'}(q,\tau)
=\frac{1}{N}\sum_k
G^{\alpha\alpha}_{\uparrow}(k+q,-\tau)\,
G^{\alpha'\alpha'}_{\downarrow}(k,\tau),
\label{eq:sus_ssh_hk_tau}
\end{equation}
with $\alpha,\alpha'\in\{a,b\}$.  
Since the Hamiltonian conserves the particle number on each sublattice, all
$G^{\alpha\alpha'}$ coincide in one dimension; hence sublattice indices may be
suppressed.  The charge and spin responses differ only by a factor of two.

The retarded Green’s function admits a spectral representation in a compact branch-sum form,
\begin{equation}
G(k,\omega)
=\sum_{i}\frac{w_k^{(i)}}{\omega+i\eta-E_k^{(i)}},
\label{eq:G_branch_compact}
\end{equation}
where $E_k^{(i)}$ denote the many-body excitation energies and $w_k^{(i)}$ the corresponding spectral weights. Substituting Eq.~\eqref{eq:G_branch_compact} into the linear response formula [Eq.~\eqref{eq:sus_ssh_hk_tau}] yields a susceptibility expression analogous to the HK model:
\begin{equation}
\chi_{\rm SSH-HK}(q,\omega)=\frac{1}{4N}\sum_{k}\sum_{i,j}W^{ij}_{k,k+q}
\frac{\Theta(-E_k^{(i)})-\Theta(-E_{k+q}^{(j)})}{\omega+i\eta+E_k^{(i)}-E_{k+q}^{(j)}},
\label{eq:chi_ssh_hk_compact}
\end{equation}
where $W^{ij}_{k,k+q}=w_k^{(i)}\,w_{k+q}^{(j)}$ are the branch-resolved transition weights. In this work, the chemical potential is fixed at $\mu=0$, such that the occupancy is determined by the sign of the excitation energies.

For filling $n=1$, five excitation branches contribute, with the entries in
$w_k^{(i)}$ ordered to correspond to those in $E_k^{(i)}$:
\begin{align}
\nonumber E_k^{(i)}&\in\{
-|\epsilon_{k}^+|,\,
|\epsilon_{k}^+|,\,
U+|\epsilon_{k}^+|,\,
U-|\epsilon_{k}^+|,\,
U+3|\epsilon_{k}^+|\},~~\\
w_k^{(i)}&=\{
\frac{1}{2},\,
\frac{1}{2},\,
\frac{1}{2},\,
\cos^2\phi_k,\,
\sin^2\phi_k
\}.
\label{eq:branches_n1}
\end{align}
For filling $n=2$, four excitation branches contribute:
\begin{align}
\nonumber E_k^{(i)}&\in\{
U+|\epsilon_{k}^+|,\,
U-|\epsilon_{k}^+|,\,
U+3|\epsilon_{k}^+|,\,
U-3|\epsilon_{k}^+|\},~~ \\
w_k^{(i)}&=\{
\frac{1}{2},\,
\cos^2\phi_k+\sin^2(\frac{\phi_k}{2}),\,
\frac{1}{2},\,
\sin^2\phi_k+\cos^2(\frac{\phi_k}{2})
\}.
\label{eq:branches_n2}
\end{align}

Inserting these expressions into Eq.~\eqref{eq:chi_ssh_hk_compact} gives the
full dynamical susceptibility of the SSH--HK model for the corresponding
filling.
\begin{figure}[h]
	\centering
		\includegraphics[scale=0.4]{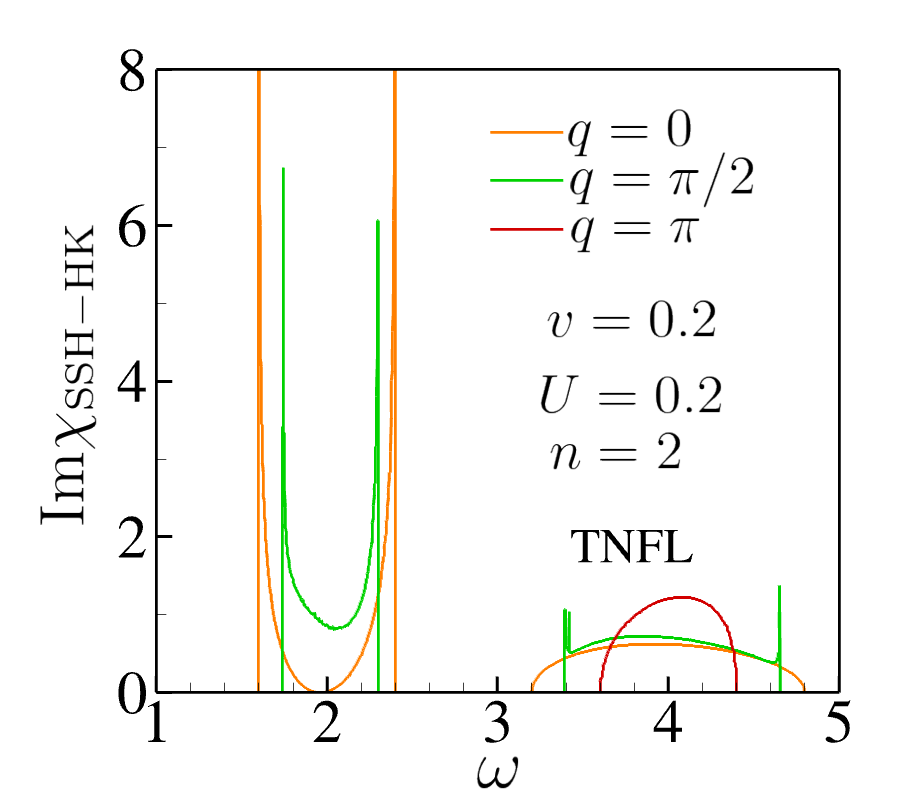}
		\includegraphics[scale=0.4]{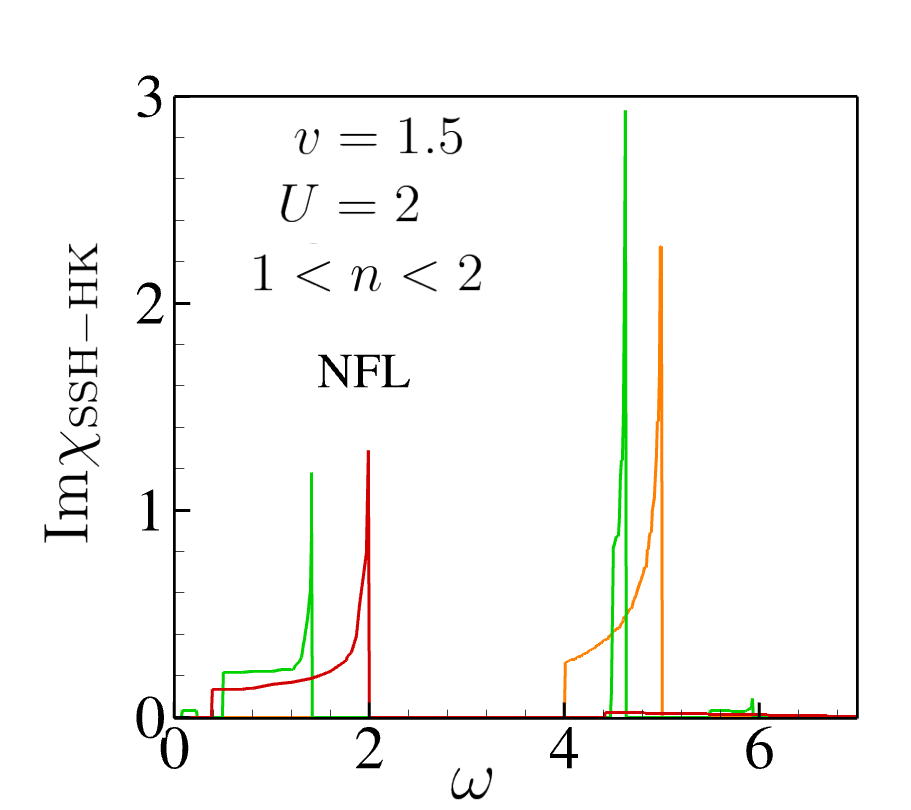}
		\includegraphics[scale=0.4]{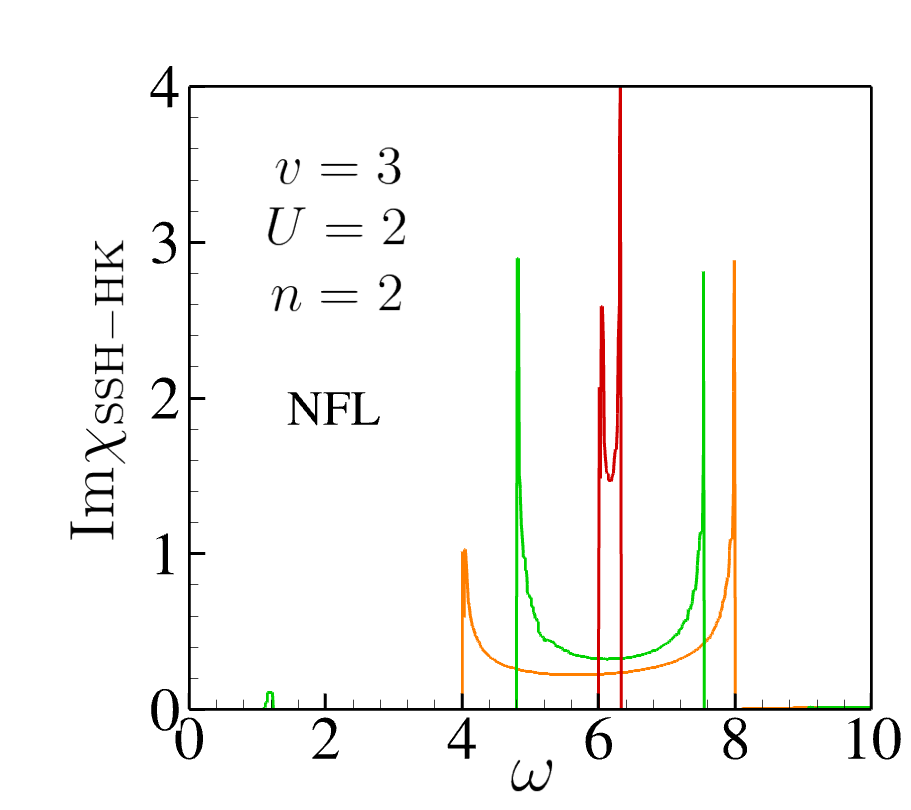}
\caption{Imaginary part of the dynamical spin susceptibility $\i\chi_{\rm SSH-HK}(q,\omega)$ as a function of frequency $\omega$ for selected wave vectors $q$. The intercell hopping amplitude is fixed at $w=1$. 
\label{fig:susceptibility}} 
	\end{figure}

In Fig.~\ref{fig:susceptibility}, the imaginary part of the spin susceptibility $\i\chi_{\rm SSH-HK}(q,\omega)$ is plotted as a function of frequency $\omega$ for several wave vectors $q$. Each peak corresponds to a $\delta$-function associated with an inter-branch excitation, with amplitude determined by the transition weight $W^{ij}_{k,k+q}$ [Eq.~\eqref{eq:chi_ssh_hk_compact}].

For $q=0$, the dynamical spin susceptibility of the SSH--HK model differs qualitatively from that of the SSH--Hubbard model. In the SSH--Hubbard case, where the response is described within the RPA framework in terms of particle-hole excitations between the SSH energy bands, intra-band processes do not contribute at $q=0$ because both the excitation energy $\epsilon^{(\nu)}_{k+q}-\epsilon^{(\nu)}_{k}$ and the occupation-factor difference [$\Theta(-\epsilon_k^{(\nu)})-\Theta(-\epsilon_{k+q}^{(\nu)})$] vanish. The remaining inter-band contributions cancel due to spin conservation, yielding a vanishing uniform susceptibility.

In contrast, the exact solution of the SSH–HK model yields several distinct branch-to-branch excitation processes, each contributing to the dynamical response. Although the pure HK model forms an NFL for $U>0$, its uniform dynamical susceptibility also vanishes. The finite value of $\i\chi(0,\omega)$ in the SSH--HK case therefore originates from the interplay between SSH bond dimerization and the HK interaction structure. The dimerized hopping mixes the sublattice degrees of freedom and redistributes spectral weight among the excitation branches [Eqs.~\eqref{eq:branches_n1} and \eqref{eq:branches_n2}], modifying the transition weights and lifting the cancellations present in the pure HK limit.

Consequently, although intra-branch excitations ($i=j$) remain forbidden at $q=0$, inter-branch processes with energy
\[
\Delta\epsilon^{ij}_{k\rightarrow k+q}=E^{(j)}_{k+q}-E^{(i)}_{k}
\]
remain allowed even in the uniform limit. These processes produce a finite imaginary part of the susceptibility, so that $\i\chi(0,\omega)$ directly reflects the combined effects of bond dimerization and the nonlocal HK interaction.

According to the ground-state phase diagram reported in Ref.~\cite{mohamadi2025emergence}, three filling regimes occur: $n=1$, $1<n<2$, and $n=2$. Their dynamical spin responses are discussed below.

For filling $n=2$ and parameters $(U,v,w)=(0.2,0.2,1)$ [Fig.~\ref{fig:susceptibility} (top)], the system lies in the TNFL phase. The largest spectral weight appears at $q=0$, where the response extends over two finite-frequency windows, indicating multiple inter-branch excitation processes. As the wave vector approaches $q=\pi$, one window closes and the remaining peak weakens.

For $(U,v,w)=(2,3,1)$ [Fig.~\ref{fig:susceptibility} (bottom)], the system remains in the NFL phase at the same filling. The susceptibility is finite within a single frequency window whose width and peak intensity depend on $q$. At $q=0$ the interval is widest but the spectral weight is smallest, while near $q=\pi$ the window narrows and the peak intensity increases, reflecting a redistribution of transition weights.

For fractional filling $1<n<2$ and parameters $(U,v,w)=(2,1.5,1)$ [Fig.~\ref{fig:susceptibility} (middle)], the system is also in the NFL phase. The susceptibility remains finite within a bounded frequency range whose structure depends on $q$. A single spectral window appears at $q=0$, while multiple windows emerge at $q=\pi/2$ and $q=\pi$, reflecting the redistribution of spectral weight encoded in the transition weights.

At filling $n=1$, the imaginary part of the susceptibility vanishes for all $q$ and $\omega$. Although the excitation branches of Eq.~\eqref{eq:branches_n1} are well-defined, none contribute to $\i\chi(q,\omega)$. In the expression for $\chi_{\rm HK}(q,\omega)$ [Eq.~\eqref{eq:chi_ssh_hk_compact}], the combination of occupation factors and branch-resolved transition weights effectively eliminates all inter-branch excitation processes: for every pair of branches $(i,j)$, either the occupation difference $\Theta(-E_k^{(i)})-\Theta(-E_{k+q}^{(j)})$ vanishes or the corresponding transition weight is zero. As a result, no finite-frequency spin response can be generated by an external magnetic field.

\section{Dynamical susceptibility and quantum Fisher information}\label{sec:qfi}

Having established the unified formalism for the dynamical susceptibility $\chi(q, \omega)$, we now consider an integrated quantity that captures the overall strength of the dynamical response. A particularly relevant choice is the quantum Fisher information (QFI), $F_Q$, which is related to the imaginary part of the susceptibility through the following sum rule \cite{hauke2016measuring}:
\begin{align}\label{eq:fq}
	F_Q(q)= \frac{4}{\pi} \int_0^\infty d\omega~\i\chi(q,\omega).
\end{align}
Beyond its role as a spectral integral, the QFI serves as a robust witness for genuinely multipartite entanglement. This identity allows us to extract entanglement information directly from the calculated response functions, bridging the gap between theoretical many-body correlations and quantifiable entanglement measures.

\begin{figure}
	\centering
	\includegraphics[scale=0.4]{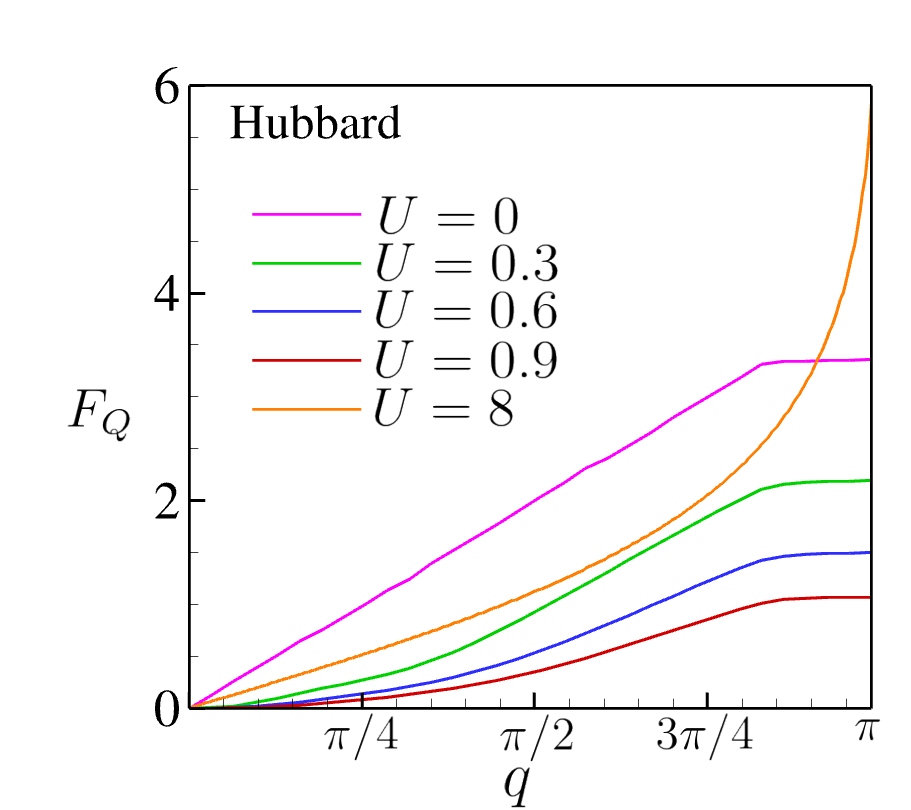}
	\includegraphics[scale=0.4]{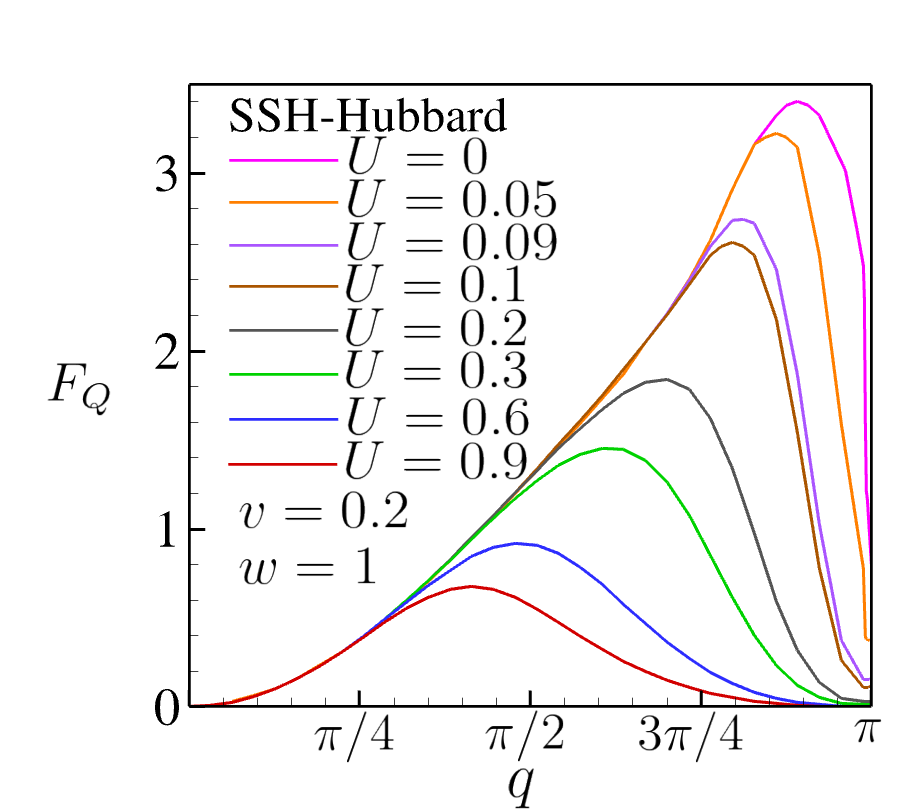}
	\includegraphics[scale=0.4]{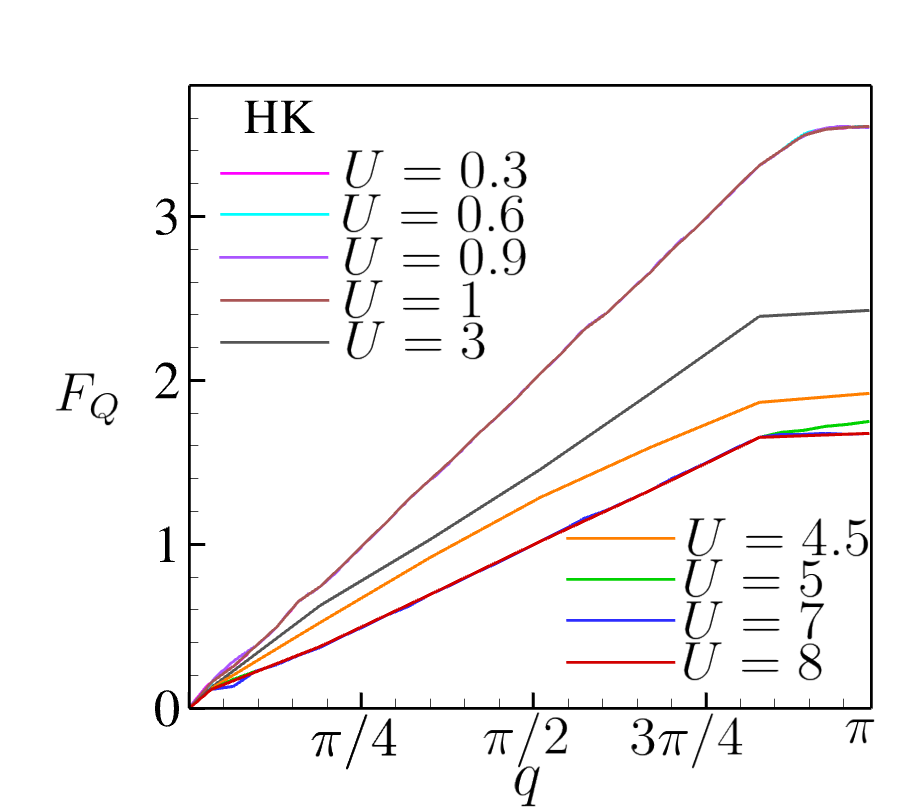}
	\includegraphics[scale=0.4]{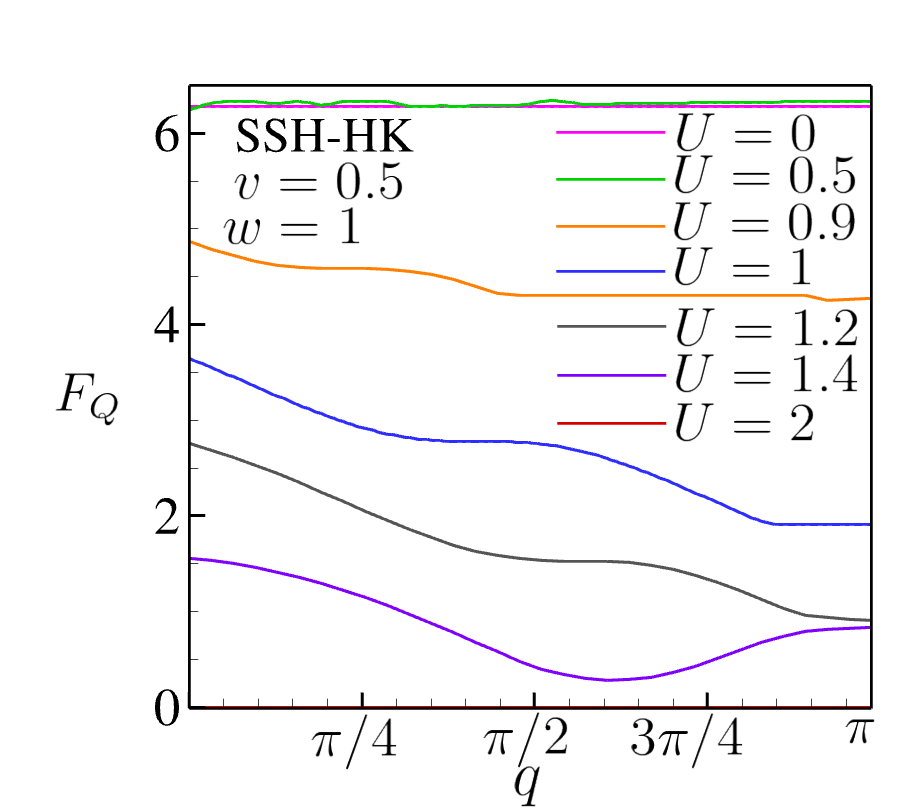}
	\caption{Quantum Fisher information as a function of $q$ for several interaction strengths. From top to bottom, the panels correspond to the Hubbard, SSH--Hubbard, HK, and SSH--HK models.} 
	\label{fig:QFI_Hubbard}
\end{figure}
In Fig.~\ref{fig:QFI_Hubbard}, we show the QFI as a function of wavevector $q$ for the four interacting models considered: Hubbard, SSH--Hubbard, HK, and SSH--HK. The distinct profiles reflect differences in spectral structure and interaction mechanisms among the models. Results for each case are discussed below.

{\bf Hubbard model.}
For the Hubbard model, $F_Q(q)$ exhibits a pronounced peak at $q=\pi$. 
At half-filling, the Fermi momentum satisfies $k_{\mathrm F}=\pi/2$, so the enhancement at $2k_{\mathrm F}=\pi$ is consistent with the usual $2k_{\mathrm F}$ correlation dominant in single‑band systems. 
Within the weak‑coupling RPA limit ($U<t$), $F_Q$ decreases with increasing $U$ for all $q\neq0$, while $F_Q(q=0)=0$. 
Although stronger interactions are commonly associated with enhanced ground‑state entanglement, this trend arises from a different mechanism: in RPA, interactions perturb the noninteracting ground state without introducing additional entanglement. They primarily redistribute spectral weight in $\i\chi(q,\omega)$, suppressing low‑energy excitations. 
Because $F_Q(q)$ integrates this spectral weight, its magnitude follows the same suppression with increasing $U$.

Beyond the weak‑coupling regime, $F_Q$ changes qualitatively. 
For large $U$ (e.g., $U=8$), the curve deviates from the RPA behavior and increases around $q=\pi$. 
In this limit, the low‑energy physics maps onto a spin‑$\tfrac{1}{2}$ Heisenberg chain where excitations are spinons rather than particle‑hole pairs. 
The QFI is then governed by the spinon continuum described by the Müller ansatz and becomes independent of $U$, since the interaction simply rescales energies via the superexchange parameter $J=4t^2/U$. 
Consequently, all sufficiently large-$U$ curves collapse onto a single universal form.

{\bf SSH--Hubbard model.}
The SSH–Hubbard model displays a qualitatively different $q$‑dependence. 
Already at $U=0$, $F_Q(q)$ differs from the single‑band Hubbard result because of the two‑band SSH structure, which introduces band‑dependent form factors in the susceptibility [Eq.~\eqref{eq:sus-SSH--Hubbard}]. 
For weak interactions ($U<1$), the QFI vanishes at $q=0$, increases monotonically with $q$, reaches a maximum, and then drops to zero near $q=\pi$. 
The peak height and position both depend on $U$: the maximum value occurs at $U=0$ near $q=\pi$, and increasing $U$ suppresses the amplitude while shifting the peak toward smaller $q$. 
This shift originates from the folded Brillouin zone of the SSH chain, which renormalizes the effective Fermi wavevector within the two‑site unit cell.

{\bf HK model.}
The HK model shows a profile similar to the Hubbard case—$F_Q(q)$ grows steadily toward $q=\pi$. 
For $U=0$, both models coincide since they share the same free‑fermion dispersion. 
For $U/t<1$, the curves nearly overlap; for $U/t>1$, $F_Q$ decreases as spectral weight is transferred to higher energies and available phase space is reduced, eventually saturating for larger $U$. 
Beyond $U/t\gtrsim5$, all curves collapse onto a single asymptotic form (Fig.~\ref{fig:QFI_Hubbard}). 
Unlike the Hubbard model, where strong \(U\) leads to enhanced entanglement in the Mott phase, the HK interaction yields nearly constant QFI in both weak‑ and strong‑coupling limits, with a minimum only at intermediate \(U\).

{\bf SSH--HK model.}
The SSH--HK model exhibits the most distinct behavior. 
At $U=0$, its QFI curve does not coincide with the SSH–Hubbard result at $U=0$, reflecting the difference in filling factor. 
In the noninteracting limit, the SSH--HK ground state lies in the $n=2$ sector, whereas the SSH--Hubbard case corresponds to half‑filling ($n=1$). 
As established in the SSH--HK ground‑state phase diagram \cite{mohamadi2025emergence}, the $n=1$ phase becomes stable for any finite interaction ($U\neq0$).

Figure~\ref{fig:QFI_Hubbard} shows $F_Q(q)$ for several interaction strengths. 
In phases with uniform filling, $F_Q$ remains constant: in the $n=1$ phase ($U=2$, red line) it vanishes, while in the $n=2$ phase ($U=0$, pink line) it takes a value of $2\pi$. 
In these uniform sectors, $F_Q$ is independent of $q$ and of all system parameters, including $U$. 
By contrast, in the fractional‑filling regime ($U=0.5$–$1.4$), $F_Q$ varies continuously between $0$ and $2\pi$ depending on $q$, reflecting the coexistence of partially occupied spectral branches and the corresponding evolution of dynamical correlations.

\begin{figure}
	\centering
	\includegraphics[scale=0.49]{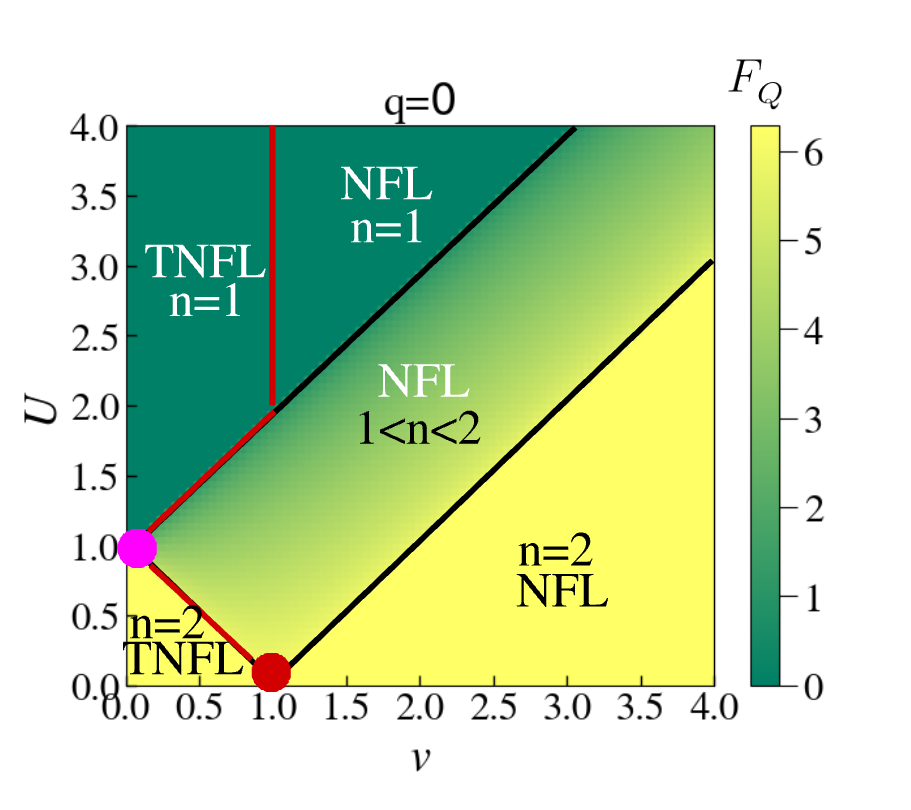}
	\includegraphics[scale=0.49]{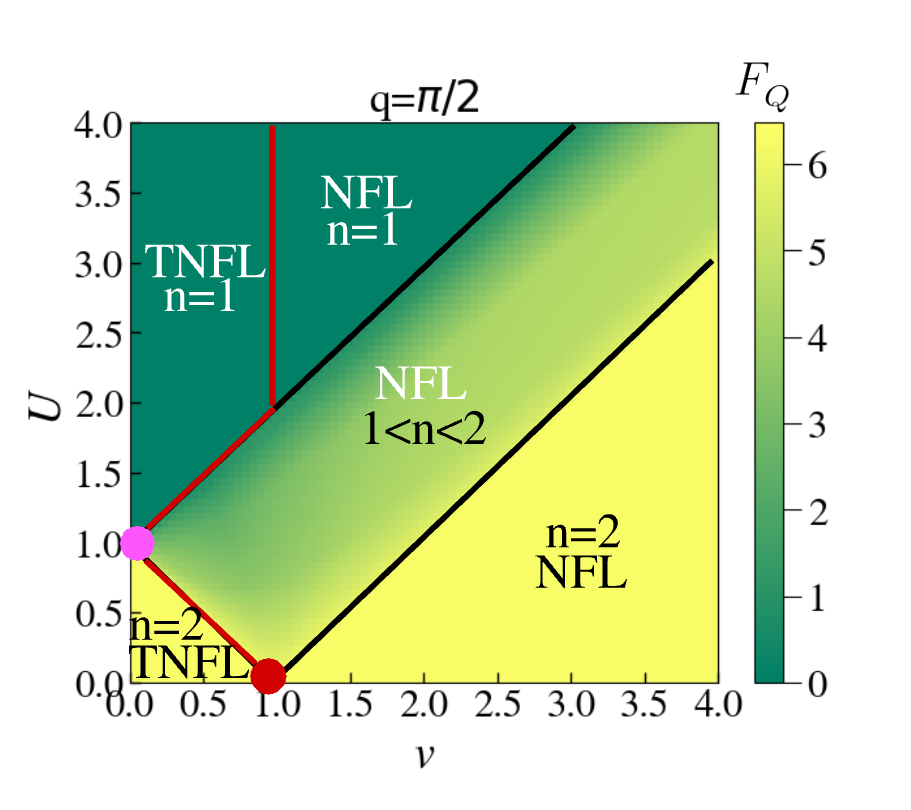}
	\includegraphics[scale=0.49]{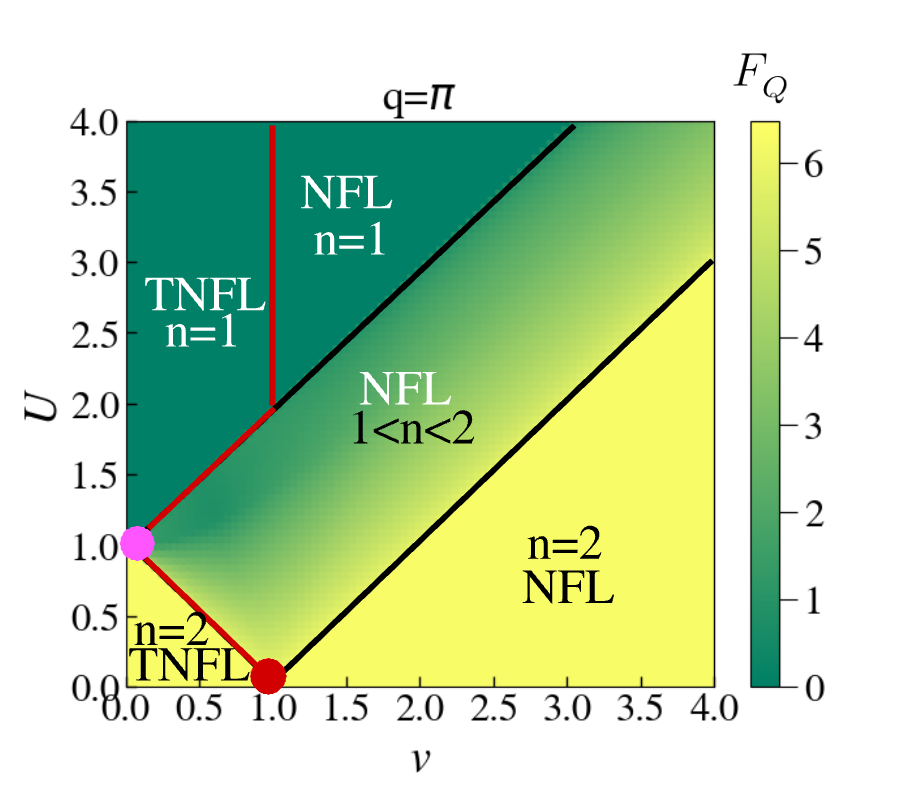}
	\caption{
		Quantum Fisher information as a function of the intracell hopping amplitude $v$ and interaction strength $U$ for $w=1$. Black lines indicate continuous phase transitions, while the pink point at $(v,U)=(0,1)$ marks a discontinuous change in the particle density $n$. Red lines and the point at $(v,U)=(1,0)$ denote the phase boundary between the trivial non-Fermi liquid (NFL) and topological non-Fermi liquid (TNFL) phases.}
    \label{fig:sshhkqfi}
\end{figure}

In Fig.~\ref{fig:sshhkqfi}, we present the QFI for the SSH–HK model as a function of interaction strength $U$ and intracell hopping $v$ for various wavevectors $q$. Black lines demarcate phase boundaries between sectors of different particle density $n$, while red lines distinguish the topological from trivial phases. The color intensity represents the magnitude of the QFI.

The QFI exhibits a distinct piecewise structure: it remains constant within uniform-filling phases, vanishing for $n=1$ and saturating at $2\pi$ for $n=2$. In these phases, the QFI is independent of $U$, $v$, and $q$. 
By contrast, in the fractional-filling phase, $F_Q$ varies continuously between $0$ and $2\pi$ and depends strongly on $q$. This arises because, in the fractional-filling sector, both the excitation energies $E_k^{(i)}$ [Eqs.~\eqref{eq:branches_n1} and \eqref{eq:branches_n2}] and the corresponding transition weights entering Eq.~\eqref{eq:chi_ssh_hk_compact} retain a nontrivial $q$-dependence, so that the inter-branch excitation processes do not cancel.

Notably, the QFI remains insensitive to the topological phase transitions occurring at $v=1$ within both the $n=1$ and $n=2$ sectors. Although these regions support distinct trivial and topological ground states, the QFI—derived from standard spin and charge susceptibilities—probes correlations that are insensitive to the underlying band topology in these uniform-filling phases. Resolving these topological transitions via dynamical response functions would require an operator specifically sensitive to the bond dimerization of the SSH chain, a direction we reserve for future investigation.

\section{Conclusion}

We investigated the dynamical spin and charge susceptibilities $\chi(q,\omega)$ in interacting lattice models ranging from the Hubbard and SS--Hubbard models (treated within the random-phase approximation) to the exactly solvable Hatsugai--Kohmoto (HK) and SSH--HK models. This allows a direct comparison between generic interaction effects and response functions that can be obtained analytically.

Within RPA, we find that the on-site repulsion suppresses $\i\chi(q,\omega)$ in both the Hubbard and SSH--Hubbard models by redistributing spectral weight and reducing the strength of low-energy particle-hole excitation processes. In the single-band Hubbard model, the response remains dominated by the conventional $2k_{\mathrm F}$ structure, yielding a pronounced enhancement near $q=\pi$ at half-filling. In the dimerized SSH--Hubbard case, the two-site unit cell qualitatively reshapes the momentum dependence and shifts the dominant response to $q<\pi$, consistent with a renormalized effective Fermi wavevector.

For the HK and SSH–HK models we derived closed-form expressions for $\chi(q,\omega)$, enabling a transparent characterization of interaction- and filling-dependent response. The HK susceptibility shows a Hubbard-like momentum profile but saturates toward a universal large-$U$ form. The SSH–HK model displays markedly different behavior, including a finite response at $q=0$ for selected fillings, reflecting the interplay of SSH dimerization and the momentum-diagonal HK interaction, and a strong evolution of the spectral weight across filling sectors.

We further analyzed the QFI, as an integrated measure of dynamical correlations and a witness of multipartite entanglement. In the SSH–HK model, $F_Q$ sharply distinguishes filling regimes: it vanishes at $n=1$, takes a constant nonzero value at $n=2$, and becomes strongly wavevector dependent for $1<n<2$. At the same time, $F_Q$ remains insensitive to the topological transition at $v=1$ within uniform-filling sectors, indicating that spin and charge response alone do not resolve the SSH dimerization topology in these regimes. Our results provide a unified response-based perspective on correlations and filling-controlled phases in these models and are directly relevant to spectroscopic probes of $\chi(q,\omega)$ in solids and cold-atom settings.


\bibliographystyle{ieeetr}
\bibliography{scibib}

\end{document}